\documentclass[pra,letterpaper,twocolumn,superscriptaddress,floatfix]{revtex4}
\usepackage{graphicx,psfrag,amsmath,amssymb,amsfonts,bbm,latexsym,color,dcolumn,bm}

\begin{document}
	
\title{Relationship between nonlinearities and thermalization in classical open systems:\\ 
The role of the interaction range}

\author{Roberto Onofrio}

\affiliation{\mbox{Dipartimento di Fisica e Astronomia ``Galileo Galilei'', Universit\`a  di Padova, 
		Via Marzolo 8, Padova 35131, Italy}}

\affiliation{\mbox{Department of Physics and Astronomy, Dartmouth College, 6127 Wilder Laboratory, 
		Hanover, New Hampshire 03755, USA}}

\author{Bala Sundaram}

\affiliation{\mbox{Department of Physics, University of Massachusetts, Boston, Massachusetts 02125, USA}}

\begin{abstract}
We discuss results on the dynamics of thermalization for a model with Gaussian interactions between 
two classical many-body systems trapped in external harmonic potentials. Previous work showed an approximate, 
power-law scaling of the interaction energy with the number of particles, with particular focus on the dependence 
of the anomalous exponent on the interaction strength.
Here we explore the role of the interaction range in determining anomalous exponents, showing that 
it is a more relevant parameter to differentiate distinct regimes of responses of the system.
More specifically, on varying the interaction range from its largest values while keeping the interaction strength constant, 
we observe a crossover from an integrable system, approximating the Caldeira-Leggett interaction term in the long range limit, 
to an intermediate interaction range in which the system manifests anomalous scaling, and finally to a regime of local 
interactions in which anomalous scaling disappears. A Fourier analysis of the interaction energy shows that nonlinearities give 
rise to an effective bath with a broad band of frequencies, even when starting with only two distinct trapping frequencies, yielding 
efficient thermalization in the intermediate regime of interaction range. 
We provide qualitative arguments, based on an analogous Fourier analysis of the standard map, supporting 
the view that anomalous scaling and features of the Fourier spectrum may be used as proxies to identify the role
of chaotic dynamics. Our work, that encompasses models developed in different contexts and unifies them in a common framework, 
may be relevant to the general understanding of the role of nonlinearities in a variety of many-body classical systems, ranging from 
plasmas to trapped atoms and ions.
\end{abstract}

\maketitle

\section{Introduction}

The transfer of energy within many-body systems is critical to the understanding of dissipative processes and equilibration
dynamics, a central problem in nonequilibrium statistical mechanics \cite{Maxwell,Boltzmann} and its broad range of 
physical applications \cite{Cercignani,Kremer}. For classical systems, various techniques to introduce coarse-graining in the dynamics 
have been implemented, resulting in either stochastic equations such as the Langevin equation, or deterministic partial differential equations 
for probabilistic quantities such as in the Fokker-Planck equation. These approaches do not extend naturally to quantum mechanical systems 
where the consideration of Hamiltonian, conservative structures and the related unitarity are strong requirements. 
For this reason, models for open systems based on Hamiltonian dynamics have been promoted since the early 1960s  
\cite{Magalinskii,Ullersma1,Ullersma2,Ullersma3,Ullersma4}, resulting in what is now known as the Caldeira-Leggett 
model \cite{Caldeira1,Caldeira2,Caldeira3}. In these models, a smaller subsystem, composed of a test particle or a set of harmonic oscillators, 
is linearly coupled to a larger subsystem, the `bath'. The latter is composed of an infinite number of harmonic 
oscillators distributed with a continuum of frequencies, and with a well-defined initial energy distribution, for instance Boltzmann-like. 
By properly imposing initial conditions on the degrees of the freedom of the bath, connections to the 
Langevin or Fokker-Planck equations are readily achieved either by considering a single realization of the test particle initial 
condition or a properly averaged one, respectively.
It is crucial in this approach that the bath is composed of a continuum of frequencies for the harmonic oscillators, as 
this ensures that energy revivals due to multiple beating between the various particles make the dynamics irreversible as 
expected for thermalization processes. 

Initially motivated by the need to understand thermalization processes in systems composed of a finite spectrum of frequencies - as usually 
occurs in trapped systems in atomic and plasma physics - we discussed thermalization in the context of a model which may be also 
considered a nonlinear generalization of the Caldeira-Leggett model~\cite{OnoSun}.
As a byproduct of the nonlinearity, we reported in~\cite{JauOnoSun1} anomalous scaling behavior for the average
total interaction energy with respect to the number of particles, for two equally balanced baths, once thermalization was approached.
Power-law scaling was observed with an exponent quantitatively close to that associated with Kolmogorov scaling
in turbulent fluid mixtures \cite{Kolmogorov}, suggestive of thermal homogenization. The model was also used to study the 
interplay of nonlinearities arising from both interaction and confining potentials~\cite{JauOnoSun2}.

\begin{figure*}[t]
\includegraphics[width=0.48\textwidth, clip=true]{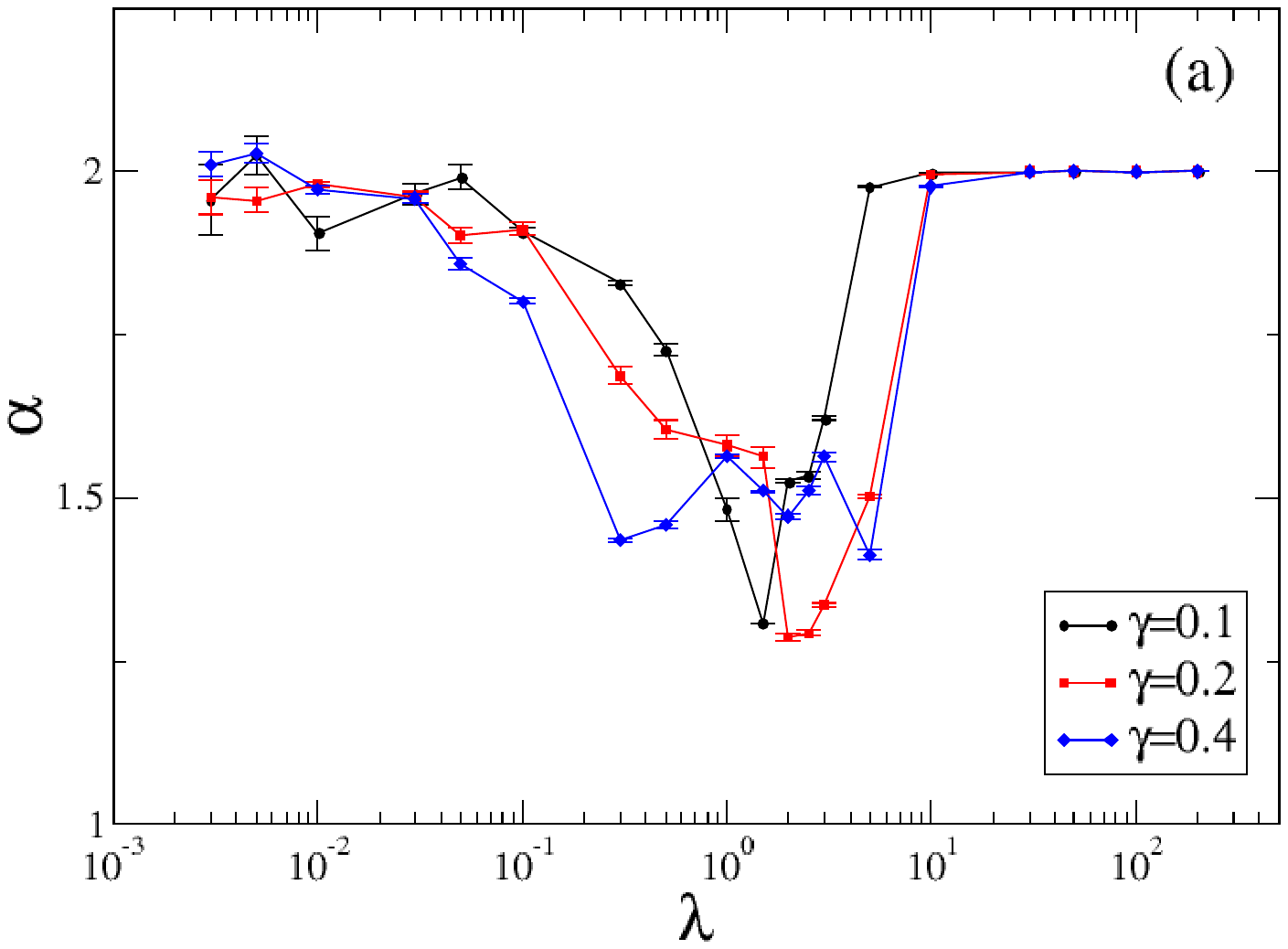}
\includegraphics[width=0.48\textwidth, clip=true]{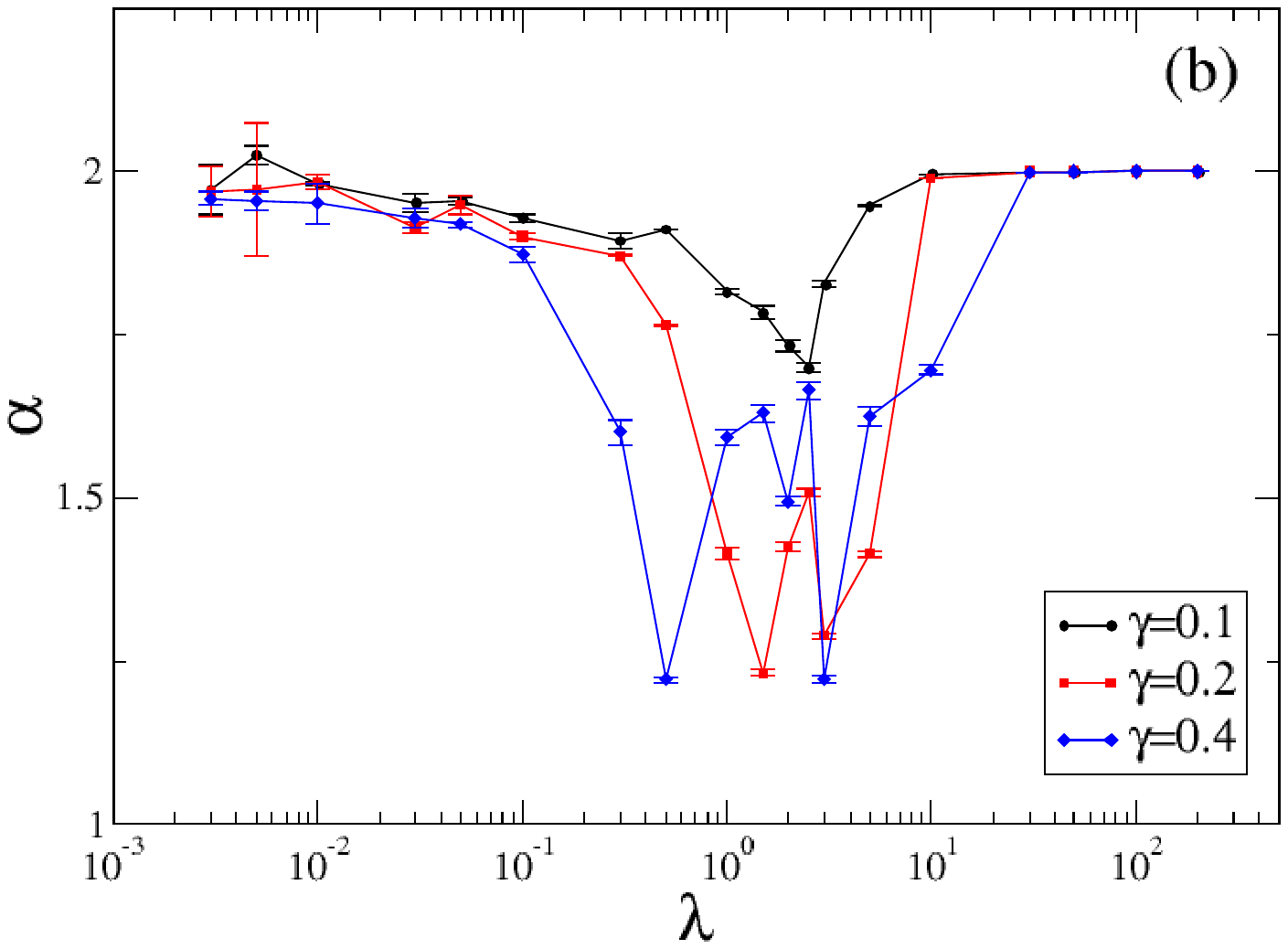}
\caption{Scaling exponent $\alpha$ versus $\lambda$, evaluated, as described in \cite{JauOnoSun3}, by fitting the interaction 
energy at long times to  power-law in the number of harmonic oscillators, $N_A=N_B=50, 100, 200, 400$.  Both short and long-range interactions do not result in anomalous scaling for the exponent $\alpha$, while at intermediate values the scaling exponents are anomalous.  The curves presented in each plot differ by the value of $\gamma$, showing a propensity to a broader and deeper anomalous region in the $\lambda$-space with increasing $\gamma$. Also, the two plots show 
a dependence on the relative values of the trapping frequency, (a) being obtained for $\omega_B/\omega_A=144/89$, (b) for the case of identical trapping frequencies,  $\omega_B/\omega_A=1.$}
\label{Fig1}
\end{figure*}

In a more recent contribution, we further explored the dependence of this anomalous scaling with the interparticle interaction 
strength~\cite{JauOnoSun3}, showing that the anomalous behavior only appears for an intermediate range of 
interaction strength values. While for simplicity we focused on one-dimensional (1D) systems, the phenomenon persists 
in higher dimensionality, differing only in the specific value of the anomalous scaling exponent. 
A perturbative, analytic approach was used to predict the values of the anomalous scaling exponents. 

In this paper we extend our analysis of the model by considering the dependence of various indicators on the range of the interaction, 
that goes from the hard-sphere model in one extreme to a lattice model resembling the Caldeira-Leggett at the other, though without the 
continuum of frequencies used in the latter. Specifically, in Sec. II, we discuss the behavior of the scaling exponent versus the 
range of the interaction, and show the morphing from the regime of rare short-range collisions amenable to the Boltzmann 
approach into the Caldeira-Leggett regime, in the limit of small interaction strength.
Anomalous exponents emerge in the  intermediate regime in which nonlinearity and chaos are expected to dominate the dynamics. 
The study also allows us to qualitatively relate the nonlinear regime to the most efficient conditions for thermalization. 
In Sec. III we discuss the dynamics of the interaction energy in Fourier space that shows the emergence of a large number of effective 
degrees of freedom in the interacting system, necessary for thermalization, despite the initial presence of just two relevant frequencies. 
It is then more manifest that nonlinearities, via frequency doubling and cascading, are responsible for the emergence of 
an effective multifrequency bath. We also identify a robust and sensitive indicator of the Fourier spectrum of the interaction energy, 
qualitatively discussing its meaning in the three different regimes. Analogous behavior is evidenced in the case of 
the Fourier analysis of the standard map for which chaotic regimes are well established.
In Sec. IV, we discuss the general applicability of the model in a number of concrete physical contexts, including the 
relationship of our findings to nonextensive statistical mechanics.

\begin{figure*}[t]
\includegraphics[width=0.95\textwidth, clip=true]{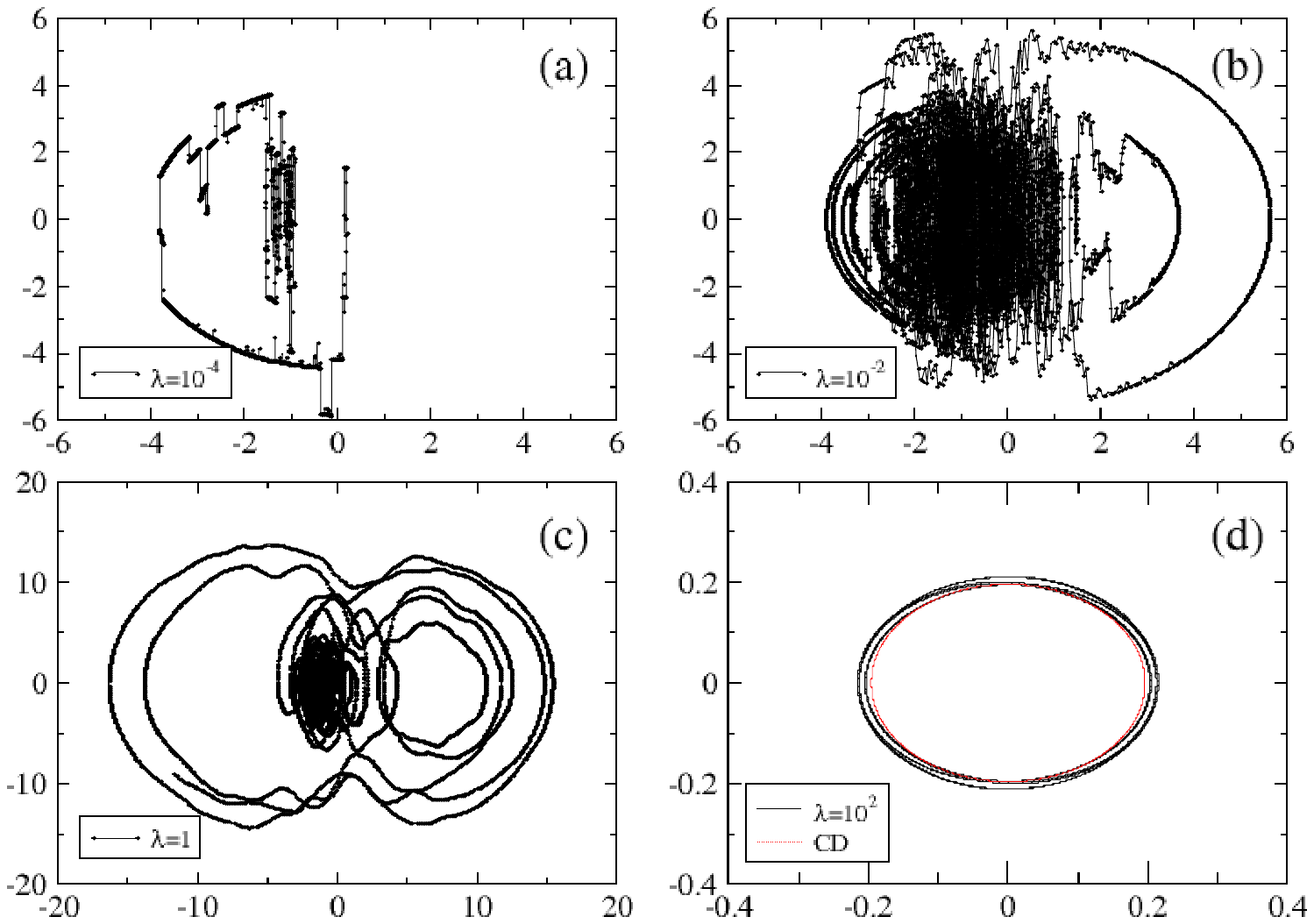}
\caption{Evolution of a particle in phase space (position in abscissa, momentum in ordinate)  for various values of 
the interaction range $\lambda$, and the same initial conditions for all particles in the system, with interaction 
strength $\gamma=1$. The initial condition in phase space for the selected particle is $(0.13455, 0.14265)$.
The presence of rare interactions between particles is evidenced in (a) corresponding to a quasi-local interaction. 
Progressive increases by two orders of magnitude of $\lambda$ correspond to plots (b) and (c) in which more frequent interactions 
appears especially in the center of the trapping potential, which the latter situation corresponding to large energy exchange, while 
(d) shows a case of very large $\lambda$ for which nonlinearities in the interaction potentials are not enough to create energy exchange. 
For comparison, in the latter case we also report, in red (innermost ellipse, thin line), the trajectory corresponds to a pure harmonic motion of the particle,  
starting from the same initial conditions but without interparticle interactions (Closed Dynamics).}
\label{Fig2}
\end{figure*}

\section{Anomalous scaling, thermalization and interaction range}

The Hamiltonian considered, inspired by a microscopic model for a meter in quantum measurement theory 
\cite{Mensky}, is~\cite{OnoSun,JauOnoSun1,JauOnoSun2}

\begin{eqnarray}
H&=&\sum_{m=1}^{N_A} \left(\frac{P_m^2}{2m_A}+\frac{1}{2} m_A \omega_A^2 Q_m^2 \right)+ \nonumber \\
& & \sum_{n=1}^{N_B} \left(\frac{p_n^2}{2m_B}+\frac{1}{2} m_B \omega_B^2 q_n^2 \right)+ \nonumber \\
& & \gamma \sum_{m=1}^{N_A} \sum_{n=1}^{N_B} \exp{\left[-\frac{\left(Q_m-q_n\right)^2}{\lambda^2}\right]},
\label{Hamilton}
\end{eqnarray}
where $(Q_m,P_m)$ and $(q_n,p_n)$ are the positions and momenta of each particle of the two species $A$ and $B$, respectively, 
where positions lie in a generic D-dimensional space, which we will assume to be D=1 in the following considerations.
The interspecies term is governed by two parameters, with $\gamma$ being the strength and $\lambda$ the range of the interaction. 
Although the interaction Hamiltonian looks rather simple, it allows for the study of a variety of situations, including equally balanced 
and unbalanced mixtures, attractive ($\gamma<0$) and repulsive ($\gamma>0$) interactions, as well as long-range 
($\lambda \rightarrow \infty$) and short-range interactions. 
Based on our former studies we only focus here on repulsive interspecies interactions, as they show more markedly anomalous 
scaling (see for example Figs. 3 and 5 in Ref.~\cite{JauOnoSun3}). In particular, for a completely unbalanced mixture 
(for instance $N_A=1$ and $N_B \rightarrow \infty$), small $\gamma$ and large interaction range, the model mimics, 
in the classical limit, the genuine Caldeira-Leggett approach used to model dissipation in open systems. 
At least in the limit of weak interactions we expect the dynamics to be determined by two dimensionless parameters 
$\gamma/(K_BT_i)$ and $\lambda/\sqrt{2 K_B T_i/(m \omega_i^2)}$, ($i$=A,B), where the two quantities in the denominators reflect the 
typical available energy and lengthscale of the corresponding thermal bath, respectively. 
The equations of motion corresponding to the Hamiltonian in Eq.~\ref{Hamilton} can be numerically integrated 
to machine precision. A plot of the time-averaged interaction energy, at the end of the times considered, 
versus number of particles, for $N_A=N_B=N$, shows power law scaling as detailed in Ref.~\cite{JauOnoSun3}.

\begin{figure*}[t]
\includegraphics[width=0.48\textwidth, clip=true]{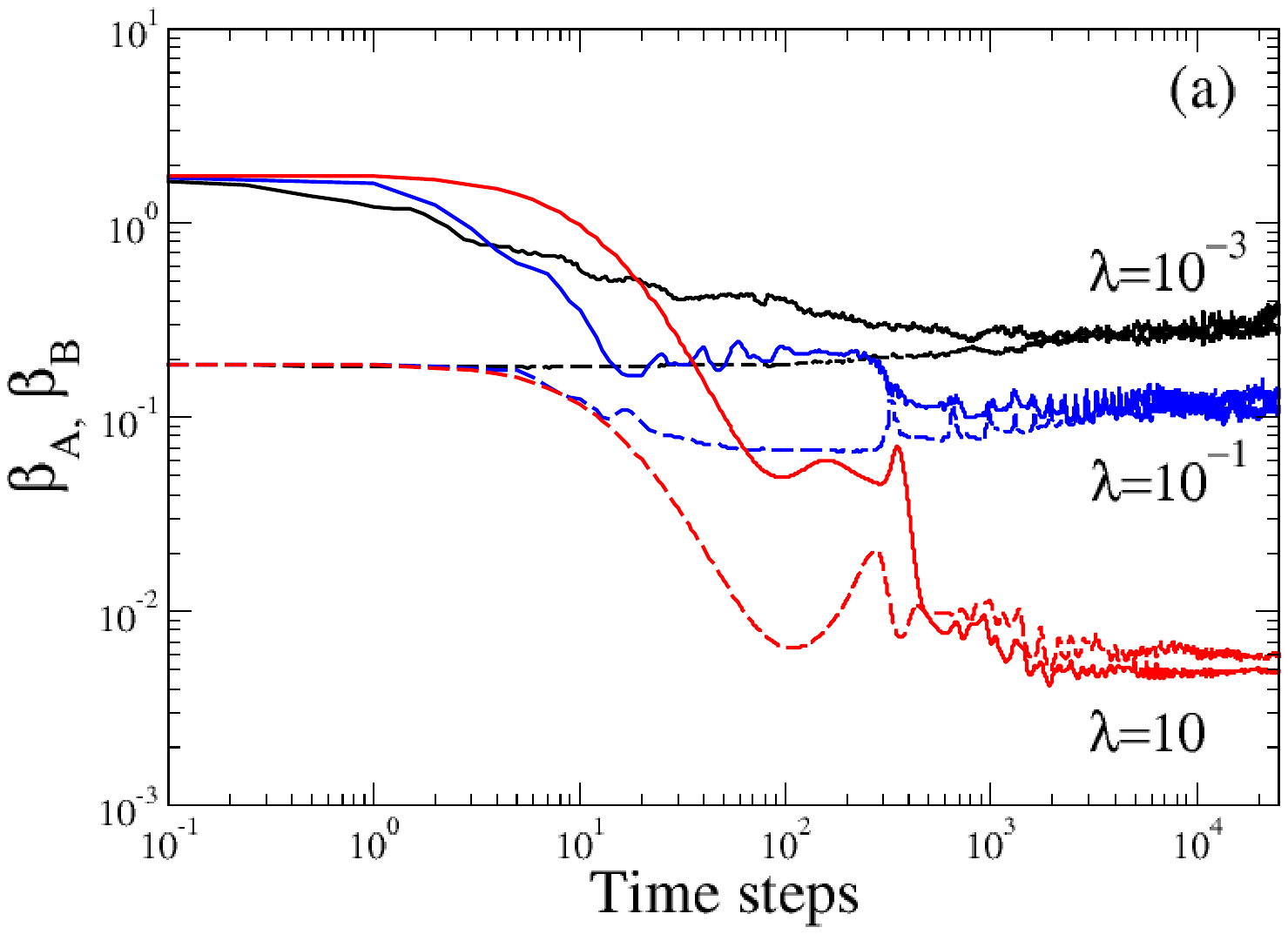}
\includegraphics[width=0.48\textwidth, clip=true]{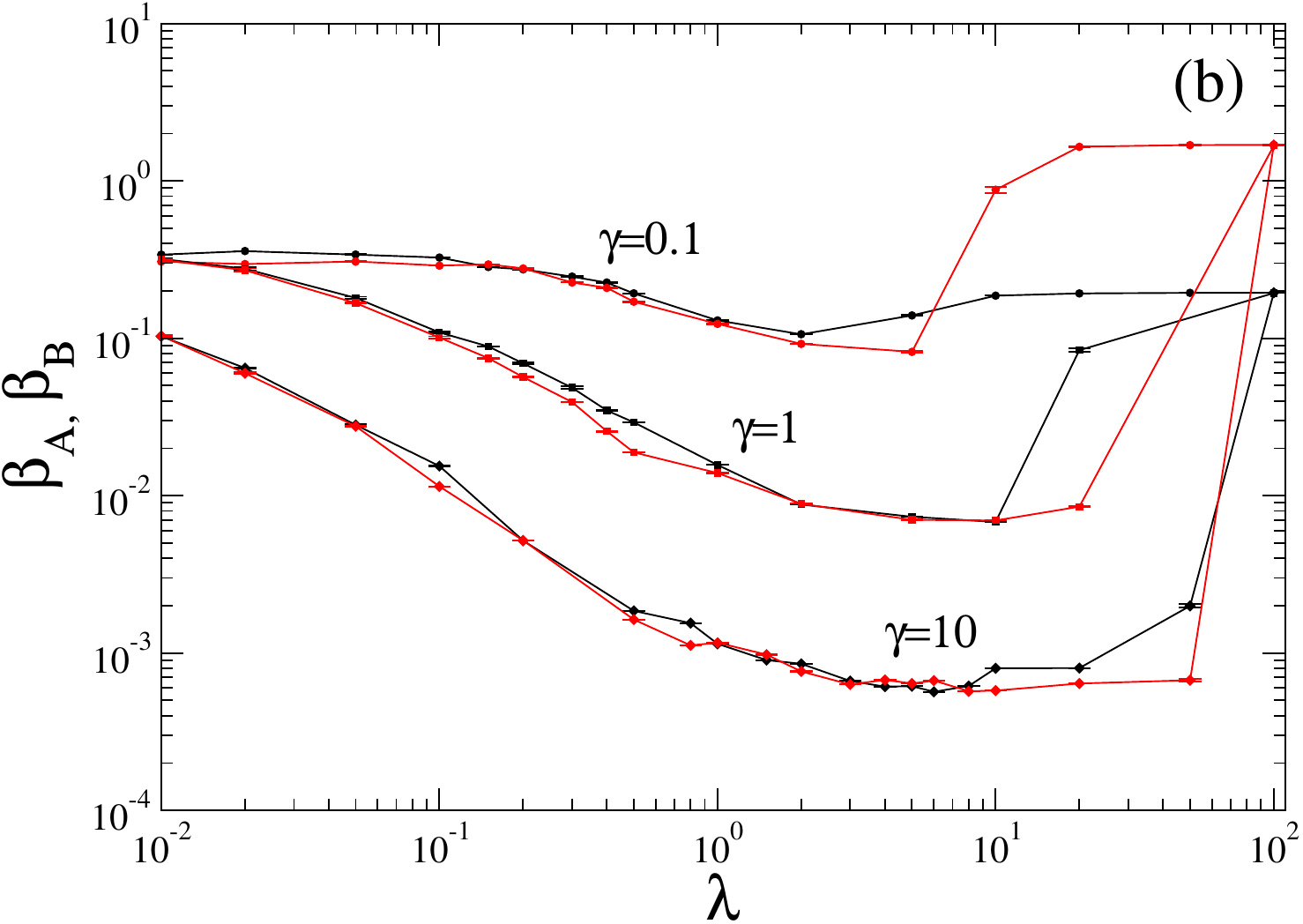}
\caption{Dependence of the dynamics of thermalization on the interaction range $\lambda$ for two systems made of 500 particles each. 
(a) Shown are the curves of the dependence of the inverse temperatures versus time for the same initial inverse temperatures 
of 0.2 (dashed lines) and 2.0 (continuous lines) in arbitrary units, and same coupling strengths $\gamma=1$. The final value of the equilibrium inverse temperature, as well as the thermalization timescales, strongly depend on $\lambda$ in the long range limit. (b) Plot of the final inverse temperatures, after $10^5$ time steps, versus the 
interaction range $\lambda$ for three different values of the interaction strength $\gamma$. This shows clearly that thermalization occurs, 
for the chosen timescale, only for small  $\lambda$ and large $\gamma$.}
\label{Fig3}
\end{figure*}

A study of the interaction energy after the onset of thermalization, defined as the time-average of  last term in the right hand side of 
Eq. \ref{Hamilton} in a time interval after thermalization 
completed
\begin{equation}
\bar{E}_{\mathrm{int}}=\gamma \sum_{m=1}^{N_A} \sum_{n=1}^{N_B} \exp{\left[-\frac{\left(Q_m-q_n\right)^2}{\lambda^2}\right]},
\end{equation}
was performed showing anomalous scaling with the particle number $N$ in each bath, $\bar{E}_{\mathrm{int}} \sim N^{\alpha}$ with 
$\alpha$ between 1 and 2, for intermediate and fixed values of $\lambda$, and varying $\gamma$ ~\cite{JauOnoSun2}. 
The results were analytically interpreted for small values of $\gamma$, in a perturbative setting where the interaction
 energy is small with respect to the total energy of the two separated systems imagined as non-interacting (corresponding to 
 the first two terms in the right hand side of Eq. \ref{Hamilton}), that allowed for the estimation of the critical exponents for various dimensionalities.
For large values of $\gamma$, we evoked an analogy to a fluid dynamics system in which strong viscosity suppresses turbulence, leading then 
to the disappearance of the anomalous exponents. In this setting, saturation regimes were identified when $\gamma$ is large and positive, corresponding 
to strong repulsion between the two systems resulting in  phase separation, and when $\gamma$ is large and negative, corresponding to clustering 
maximizing all possible interactions among all particles of the two systems.

\begin{figure*}[t]
\includegraphics[width=0.49\textwidth, clip=true]{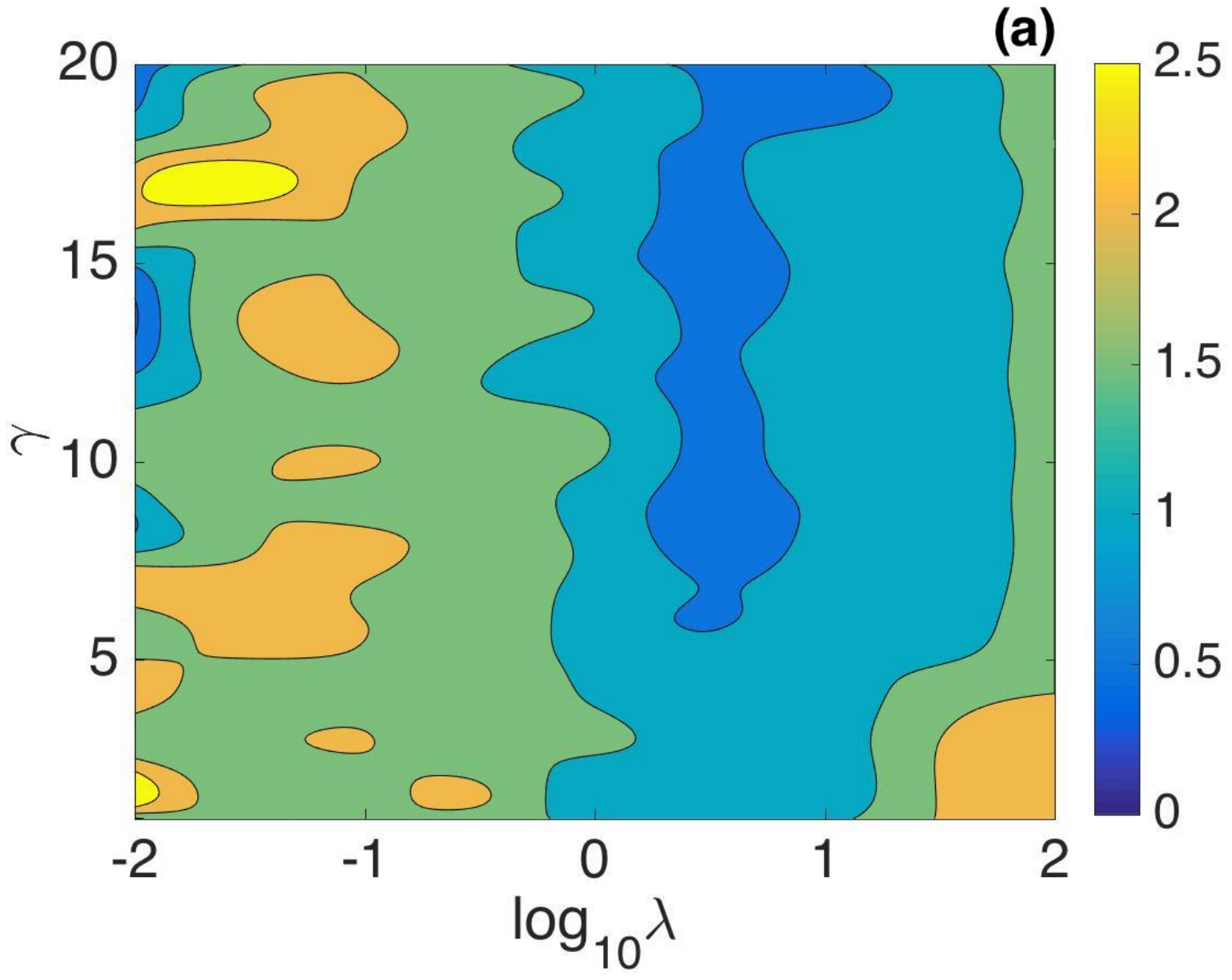}
\includegraphics[width=0.49\textwidth, clip=true]{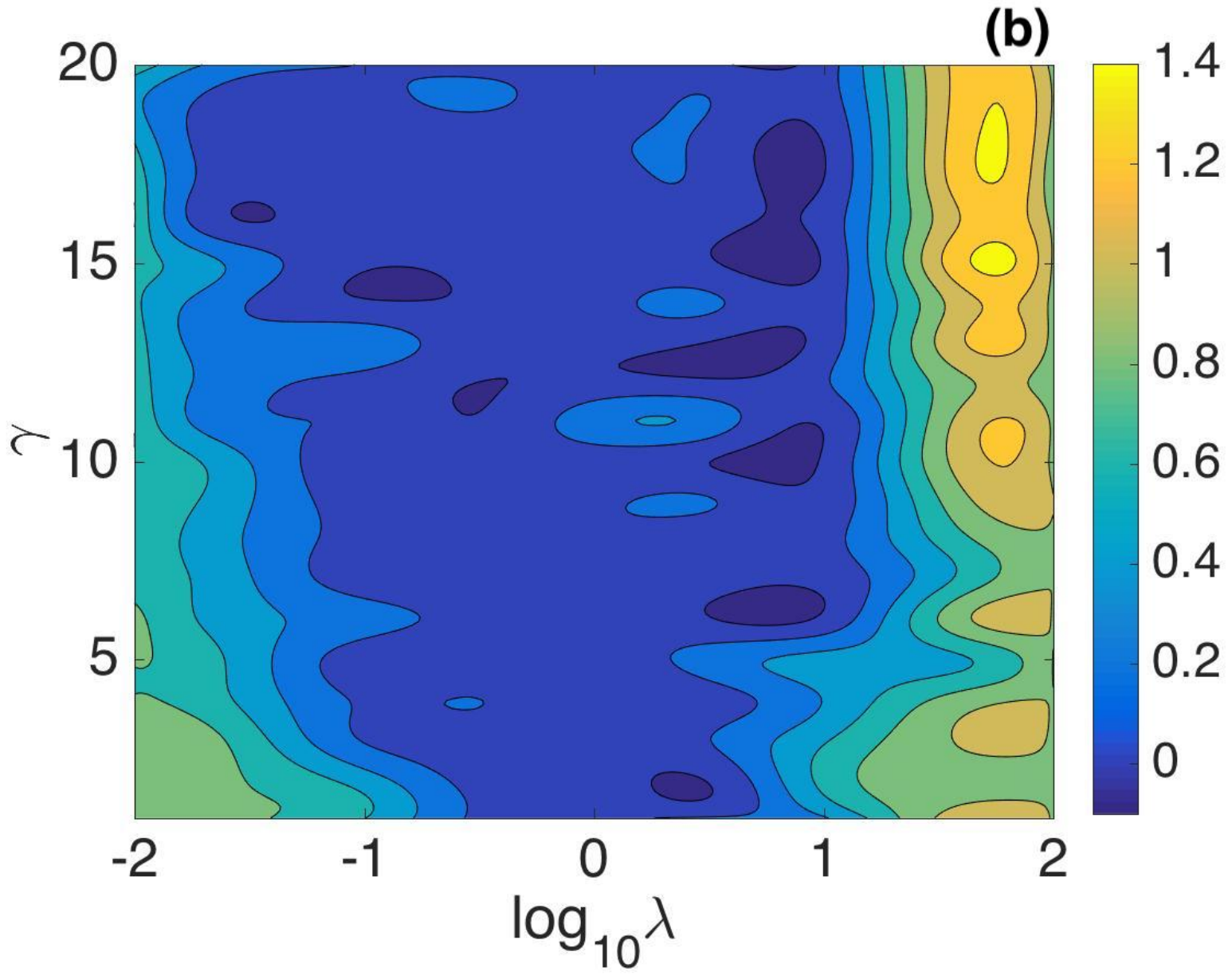}
\includegraphics[width=0.49\textwidth, clip=true]{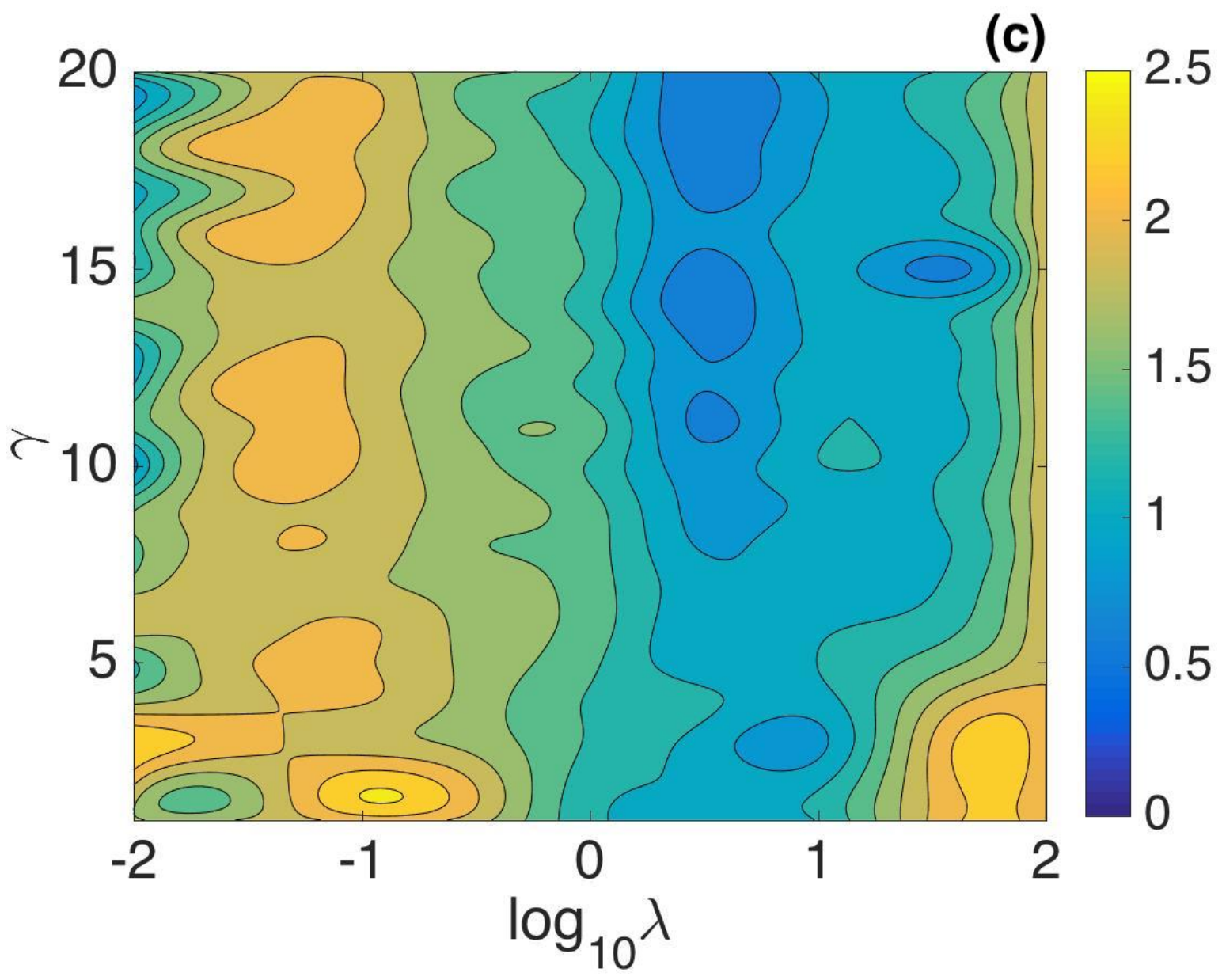}
\includegraphics[width=0.49\textwidth, clip=true]{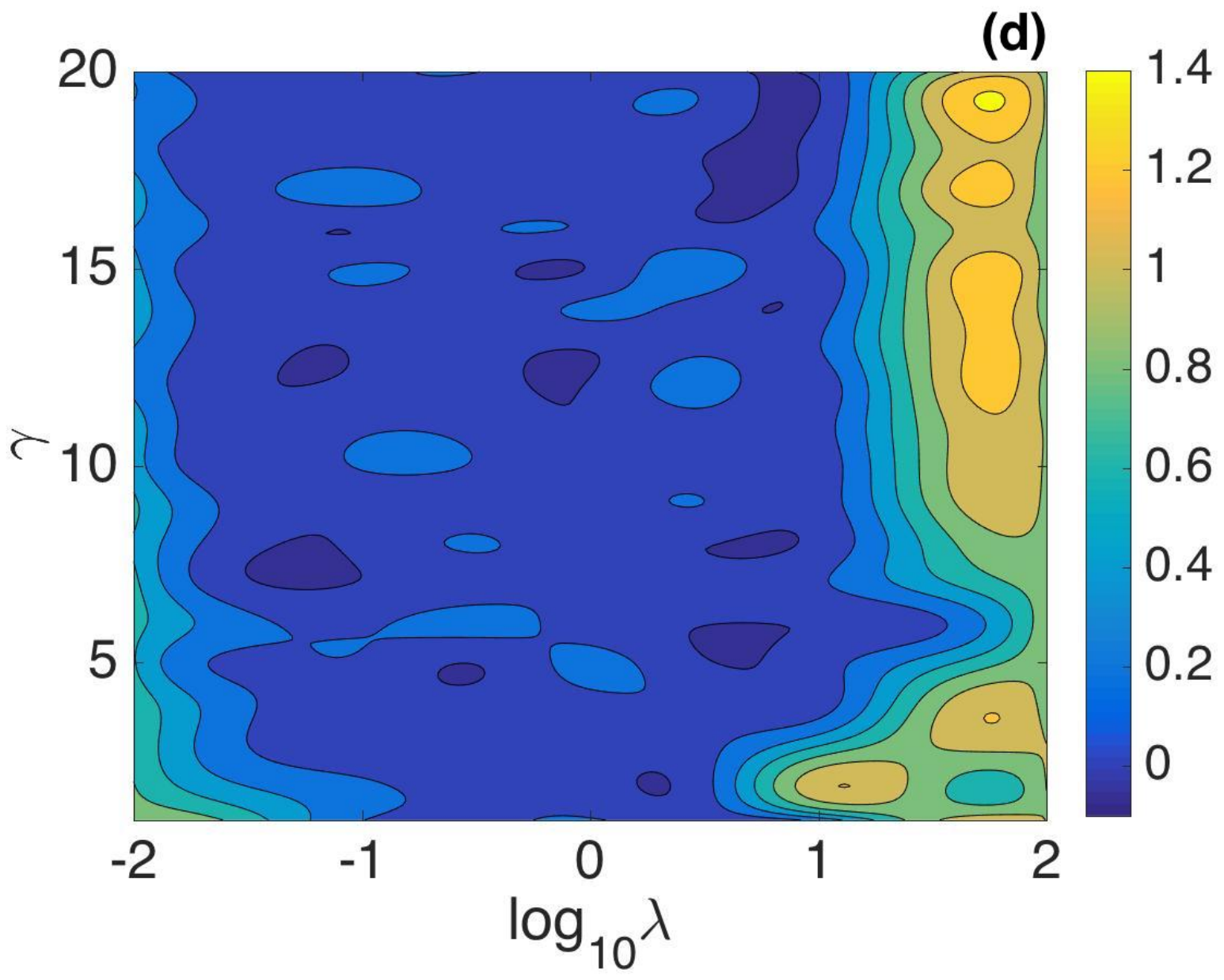}
\caption{Contour plots for the scaling exponent $\alpha$ (left plots, a and c) and the thermalization parameter $\bar{\epsilon}(t_f)$ (right plots, b and d) in the $\lambda-\gamma$ plane. 
The plots are taken after evolving the system for $10^5$ time steps (top plots) and for $10^6$ time steps (bottom plots). 
Notice the consistent behavior of the scaling exponent near 2 in the right-bottom part of the related plot, where the Caldeira-Leggett approximation is expected to hold. 
The initial inverse temperatures are 2.0 and 0.2, resulting in an initial temperature contrast of $\epsilon(0)=18/11 \simeq 1.636$.}
\label{Fig4}
\end{figure*}

In this study, we complement those findings with similarly intriguing results by assessing the dependence of the interaction energy on the interaction range $\lambda$. We use the same protocols as in~\cite{JauOnoSun2} for evaluating the scaling exponent $\alpha$, augmented, whenever sufficient, such as in the  contour plots shown later in Fig.~\ref{Fig4},  by a fast procedure for semiqualitative studies of broad parameter ranges.

In Fig.~\ref{Fig1}, we show the dependence of the scaling exponent $\alpha$ on the interaction range $\lambda$ for different values of $\gamma$. 
There is no anomalous scaling for small and large $\lambda$, where we get $\alpha=2$ with minimal error bars, especially in the Caldeira-Leggett limit. 
By contrast, in the intermediate region of $\lambda$, $\alpha$ decreases reaching a minimum of about 1.5. 
We have also explored a possible scaling of the curve with $\gamma/\lambda^2$, as the interaction term in the Hamilton equations 
(see for instance Eqs. 7,9 in \cite{OnoSun}) suggests that, at least when the interaction term dominates the dynamics, the motion is ruled by 
$\gamma/\lambda^2$. The analysis seems inconclusive in determining the scaling, and requires further exploration with other indicators 
as discussed in Section III. Even without a precisely defined scaling, it looks clear that the interaction term becomes less important as $\lambda$ increases, 
as confirmed by analyzing the full Caldeira-Leggett limit realized as $\lambda \rightarrow \infty$ and $\gamma$ small enough.

In order to emphasize the difference between the various cases even at the level of single-particle dynamics,
in Fig.~\ref{Fig2} we show the phase space dynamics for a single particle in one of the two systems, for various values of $\lambda$, spanning six orders of magnitude. 
This direct analysis in phase space corroborates the former figure. At small $\lambda$, interactions are quasi-local and therefore rare, which implies sudden changes of 
energy as observed in Fig. 2(a). Increasing $\lambda$ results in more interactions especially in the center of the trap where the particle density is higher [Fig. 2(b)], 
resulting in increased energy exchange [Fig. 2(c)]. Finally, in the Caldeira-Leggett limit [Fig. 2(d)] the particle behaves nearly as an isolated system, with no 
effective interparticle interactions. For comparison, the trajectory of isolated particle (Closed Dynamics, CD) is also shown. 

The interaction range also has a strong influence on thermalization, resulting from two competing factors. A larger $\lambda$ results in more interactions with surrounding particles at any given time. However, as mentioned before, the energy exchange, which is large when $\lambda$ is smaller than the average inter-particle distance, is suppressed as $\lambda$ grows, with a suggested scaling as $\gamma/\lambda^2$. Therefore we expect the final inverse temperature to 
initially decrease with increasing $\lambda$, followed by a return to the initial temperatures when the two systems are barely interacting in the Caldeira-Leggett limit. This dynamics is shown in Fig. \ref{Fig3}. In the left panel, we show the instantaneous inverse temperature of the two systems, $\beta_A$ and $\beta_B$, versus time for different $\lambda$, and the same value of $\gamma$. 
The instantaneous temperature is defined, as detailed in \cite{JauOnoSun1}, by averaging the energy and its square 
on the ensemble at each time step, and evaluating the energy variance, such that the inverse temperature is
\begin{equation} 
\beta=\sqrt{D}/\sigma_E,
\label{InverseT}
\end{equation}
with D the spatial dimension and $\sigma_E= (\langle E^2 \rangle - \langle E \rangle^2)^{1/2}$. Equation \ref{InverseT} holds for a Maxwell-Boltzmann 
distribution, which is a satisfactory description for most parameter cases analyzed in this paper,  apart from the extreme cases of very large $\gamma$ discussed 
in Fig. \ref{Fig4}.
Larger $\lambda$ values result in longer thermalization times and results in lower final inverse temperatures. Larger interaction ranges imply more entities responsible for thermalization, so it is intuitive that thermalization times are consequently increased. It is also reasonable that thermalization occurs at higher final temperatures, as the initial interaction energy gets larger with increasing $\lambda$. Therefore there is an increasing ``latent heat'' to be distributed among the particles in the systems.  
In the right panel, we plot the inverse temperatures after $10^5$ time steps as a function of $\lambda$, for three values of $\gamma$. 
This plot complements the inferences from the other panel, again showing that at the larger values of $\lambda$ 
there is no thermalization at least on the  timescales considered. By contrast, at small $\lambda$ the two systems track each other 
and have a common inverse temperature, that becomes  progressively lower with increasing $\lambda$, until a $\gamma$ dependent threshold value of $\lambda$ is reached, beyond which the two systems do not thermalize. We note that for 
the $\gamma=0.1$ case, the initial interaction energy, for $\lambda \leq 1$, is sufficiently small that the equilibrium temperature still lies between the initial temperatures of the two baths. For the cases of larger $\gamma$, this occurs only for a very limited range of $\lambda$ values, due to the more exothermic nature of the corresponding dynamics.

The general trend of the thermalization can be evidenced by studying the anomalous exponent $\alpha$ and the onset of thermalization versus $\lambda$ and $\gamma$. Since we were interested in a semiquantitative analysis in this case, we decided to implement a faster procedure for the estimation of $\alpha$. Instead of determining the interaction energy for five values of $N$ as in \cite{JauOnoSun2}, we have assumed a power-law dependence, consistent with 
all data collected so far, and determined $\alpha$ by comparing the time-averaged interaction energies for only two values of $N$, $N_1$ and $N_2$. For the power-law dependence we expect 

\begin{equation}
\frac{\bar{E}_{\mathrm{int}}(N_2)}{\bar{E}_{\mathrm{int}}(N_1)}=\left(\frac{N_2}{N_1}\right)^{\alpha},
\end{equation}
which immediately allows for the determination of $\alpha$. In practice, to speed up the data taking we have opted for $N_2=100$, $N_1=50$, which implies 

\begin{equation}
\alpha=\log_2 \left(\frac{\bar{E}_{\mathrm{int}}(100)}{\bar{E}_{\mathrm{int}}(50)}\right). 
\end{equation}

For the analysis of thermalization, we have considered the parameter introduced in \cite{JauOnoSun1}, {\it i.e.}

\begin{equation}
\epsilon=2\frac{|\beta_A-\beta_B|}{\beta_A+\beta_B},
\end{equation}
which is evaluated at each time step. We then consider the ratio of the average $\epsilon$ at final time, with
 an averaging window of $10^3$ time steps, and the $\epsilon$ at initial time, $\bar{\epsilon}(t_f)/\epsilon(0)$.

The outcome of this analysis is shown in Fig.~\ref{Fig4}. The left column shows contour plots for $\alpha$ while on the right, contour plots 
are shown for $\bar{\epsilon}(t_f)/\epsilon(0)$, all as a function of $\lambda$ in abscissa, and $\gamma$ in ordinate. 
The top plots are relative to the case of $10^5$ time steps, with both the interaction energies and the inverse temperatures determined 
by averaging within a time window of $10^3$ times steps. The lower plots are instead evaluated for $10^6$ time steps, to check for possible 
further gains in the thermalization process. Indeed, as expected, thermalization occurs for a broader range of values in the $\lambda-\gamma$ 
plane with a tenfold increase in the evolution time. Also worth noticing is the Caldeira-Leggett region, in the lower-right corner of each contour plot. 
In this case, as well as for all the band of values on the left side, the scaling exponent is close to 2, as expected for long range interactions involving 
all possible particle pairs. Notice also that at large $\lambda$ and  $\gamma$ the scaling exponent is anomalous. On the left plots for the thermalization, 
the less thermalizing regions are obtained for large $\lambda$, although the extent of thermalization (or lack thereof) depends on the values of $\gamma$. 
It is also evident that regions of anomalous exponents and thermalization strongly overlap, and for the chosen parameters they correspond to $\lambda$ in 
the 1-10 range.

The contour plots show a correlation between the onset of thermalization at finite times and the presence of anomalous scaling in the interaction energy with particle number. It should be noted, however, that the determination of the inverse temperatures is based on the assumption of a Boltzmann distribution as described in detail in \cite{JauOnoSun1}, which implicitly assumes a weak interaction between the two systems. In the non-perturbative regime corresponding to large $\gamma$ we do not expect this assumption to hold, and therefore the results in this regime, realized in our case if $\gamma \gtrsim 10$, are purely indicative of the qualitative behavior, requiring both verification of the energy distribution as well as the introduction of new parameters better suited for the description of these strongly coupled  systems. Also, the fact that thermalization occurs faster in the regime where anomalous scaling occurs is only indirect evidence for the possible role of chaotic dynamics. We return to this issue later with an assessment in Fourier space, by contrasting signatures in our model with those seen when considering textbook paradigm that exhibits a transition to chaotic dynamics, the standard map~\cite{LichtLieb}. For now, to better understand the relationship between thermalization and nonlinearity, we consider a different approach that provides a clear indication of the thermalization mechanism while allowing the association of a regime of chaotic dynamics with optimal thermalization.

\begin{figure*}[t]
\includegraphics[width=0.48\textwidth, clip=true]{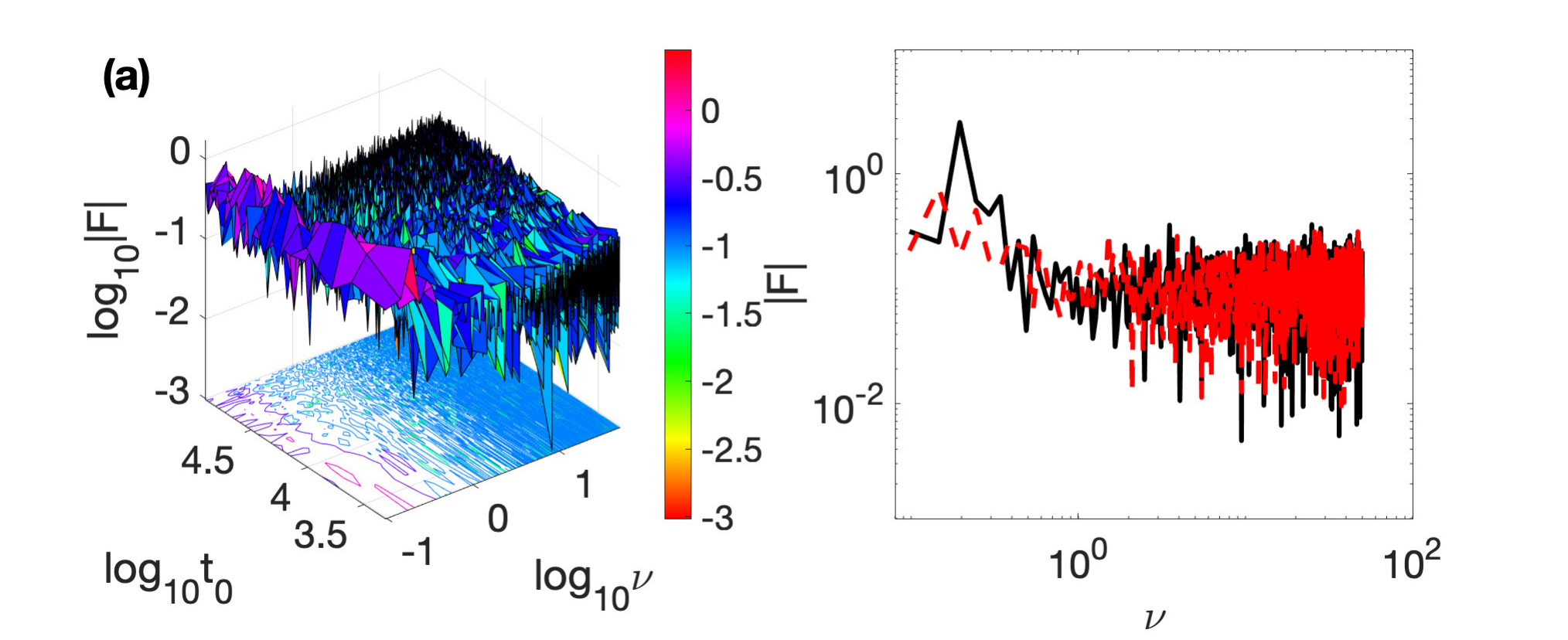}
\includegraphics[width=0.48\textwidth, clip=true]{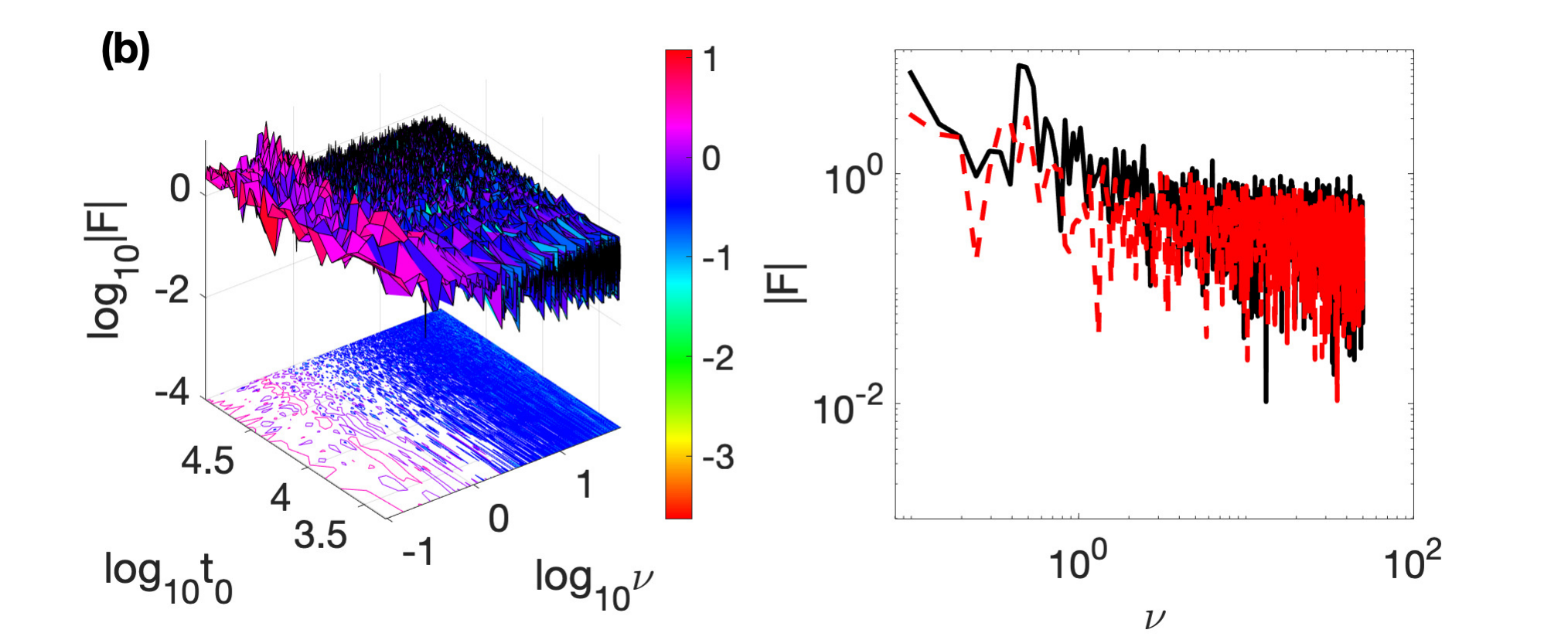}
\includegraphics[width=0.48\textwidth, clip=true]{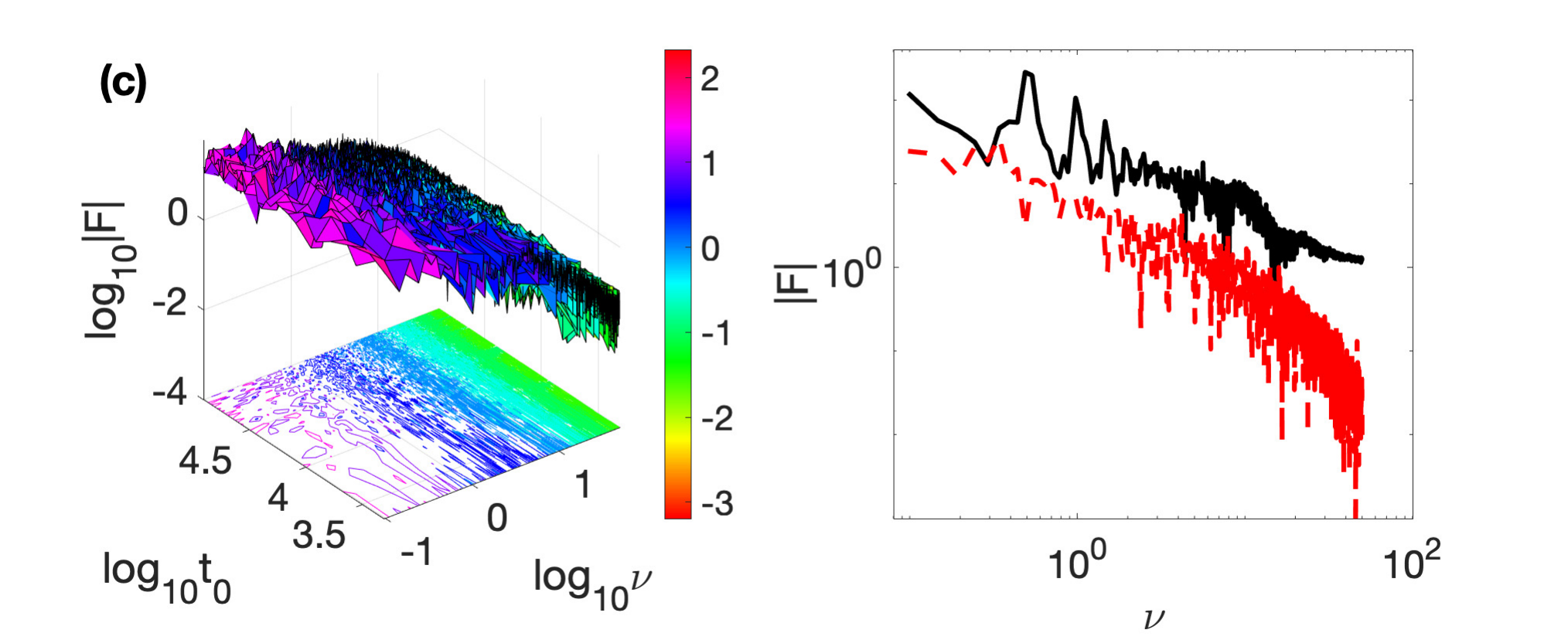}
\includegraphics[width=0.48\textwidth, clip=true]{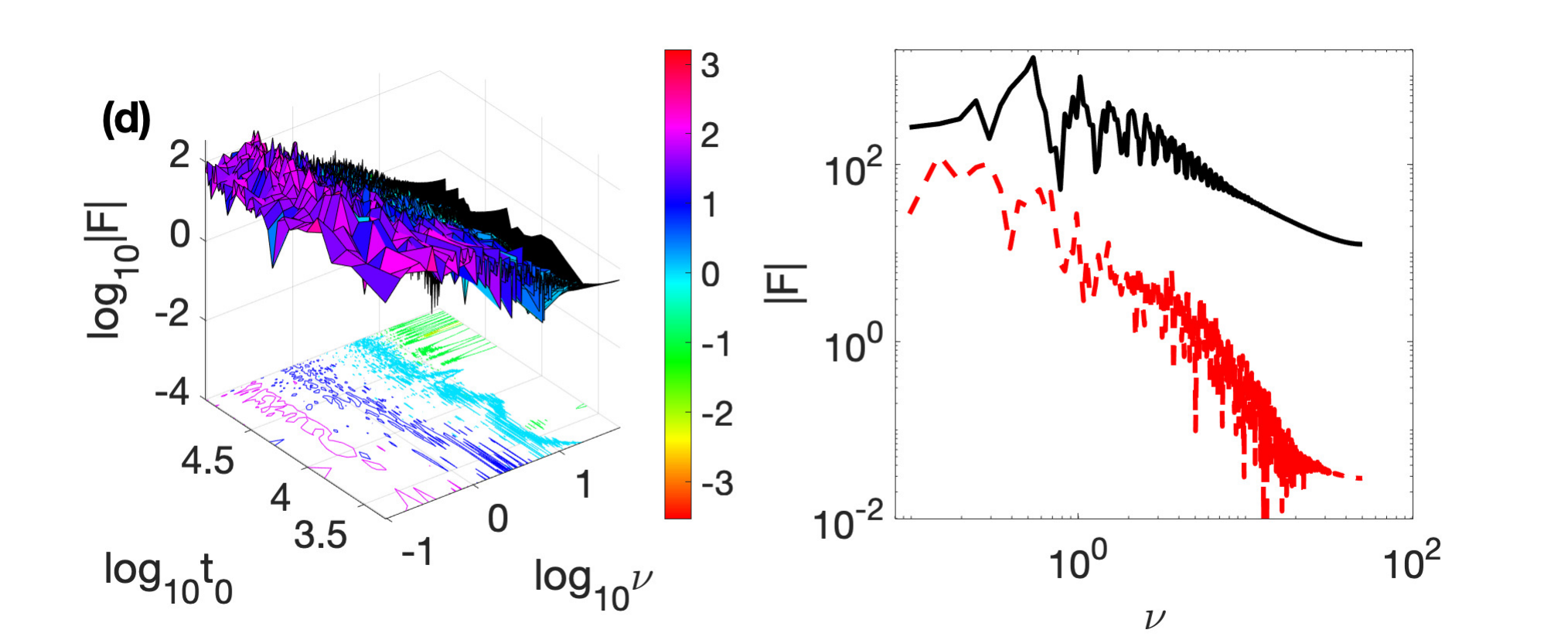}
\includegraphics[width=0.48\textwidth, clip=true]{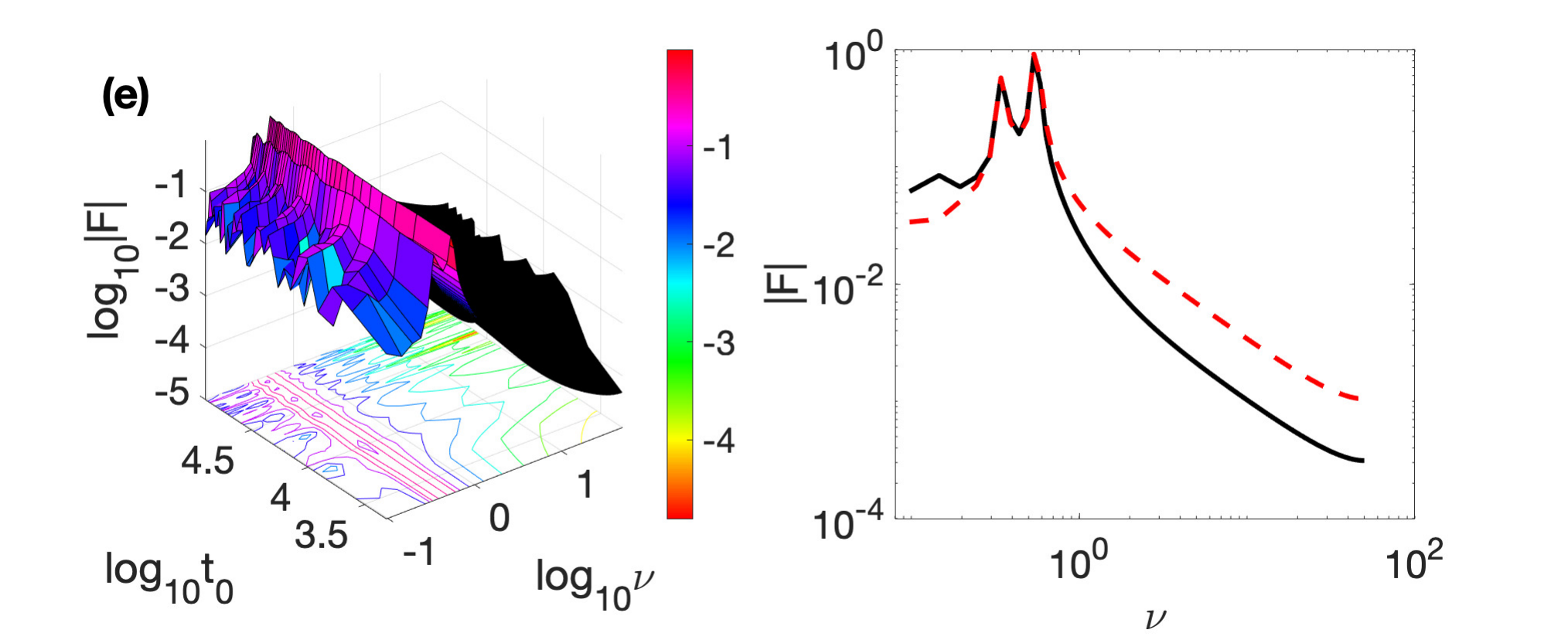}
\includegraphics[width=0.48\textwidth, clip=true]{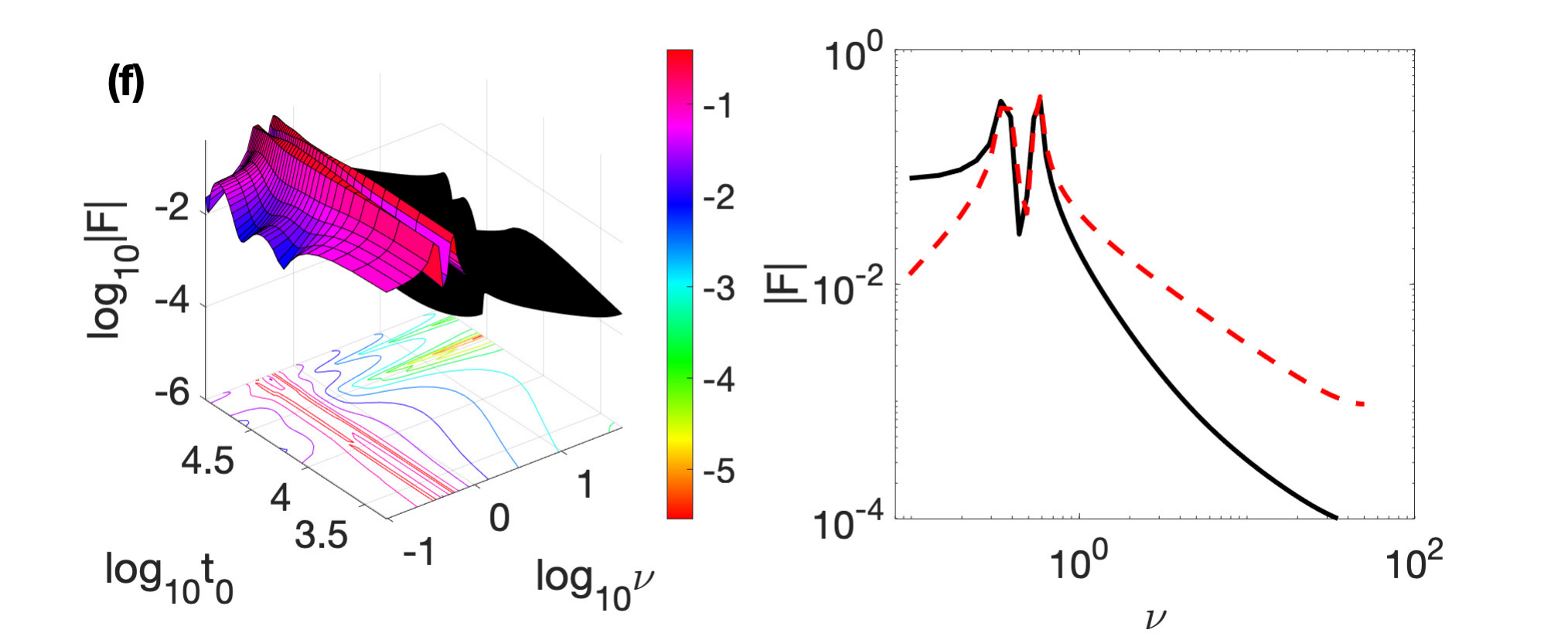}
\caption{Fourier analysis of the thermalization dynamics for $N=400$ and $\gamma=1$. 
The three-dimensional cascade plots on the left of each panel show the evolution of the FFT spectrum of the interaction energy versus time. 
The FFT is constructed by considering time series samples of 2,048 time steps.  The plots on the right of each panel show the FFT at two 
specific time windows, the initial segment (first 2,048 time steps, continuous black lines), and at the end of the full time period considered 
(last 2,048 time steps, dashed red lines). The cases shown are for (a) $\lambda=10^{-3}$, (b) $10^{-2}$, (c) $10^{-1}$, (d) $1$, (e) $10^{2}$ 
while, for contrast  to the case e),   the Fourier spectrum of the interaction energy when the particles of each bath move harmonically with equal 
amplitudes at the frequencies of $2 \omega_A/2\pi$ and $2 \omega_B/2\pi$ is shown in (f). Panel (e) corresponds to a case where the interaction term 
closely resembles the Caldeira-Leggett form, for which we do not expect the composite system to thermalize.}
\label{Fig5}
\end{figure*}

\section{Fourier analysis and emergence of effective baths}

The initial motivation for considering the generalized interaction term was the issue of how thermalization proceeds in the instance
where there is no overlap between the frequencies associated with the bath and those with the system to be thermalized. 
Earlier work had shown that, under these conditions, thermalization does not occur when the inter-species interaction is of the Caldeira-Leggett form~\cite{Smith}. 
However, the experimental technique of  sympathetic cooling suggests otherwise and this is what motivated consideration of a more general, 
nonlinear interaction term as introduced in \cite{OnoSun}. Thermalization may be viewed as akin to homogenization due to mixing in fluids, 
with exponents that reflect how well-mixed or, equivalently, thermalized the two component system has become. In the study of fluid mixing, 
the role of the dynamics (often chaotic) in producing interactions on multiple scales is well-established. This is often reflected in the spatial 
or temporal frequency spectrum associated with the dynamics. With this analogy in mind, it seems reasonable to expect similar behavior 
to be manifest in our study of thermalization. In particular, the intrinsic nonlinearity of the interaction term implies the generation of additional 
frequencies in the motion on top of the pre-existing two due to the harmonic trapping. This can be interpreted as a dynamical realization of 
an effective bath where frequencies do, indeed, overlap making thermalization possible.
   
In order to explore this possibility, it is natural to consider the time evolution of the interaction energy $\langle E_{\mathrm{int}}(t) \rangle$ which 
clearly displays increasingly quasi-periodic to aperiodic behavior with time. We consider a long time series where, by the end, thermalization has 
either occurred or there are clear signs that it will not even at very long times. We illustrate our results, in Fig.~\ref{Fig5}, for cases where each 
subsystem consists of $400$ particles. We consider a set of cases where we hold the interaction strength $\gamma$ fixed and vary the range 
of the interaction potential $\lambda$. The large $\lambda$ limit closely approximates the Caldeira-Leggett dynamics. 
In each case, the dynamics is run for long times ($10^5$ steps) with a sampling time of $0.01$ that corresponds to a Nyquist frequency 
of $50$ (times and frequencies in arbitrary units). The series is then broken up into time segments of $2^{11}=2048$ time steps and 
the Fast Fourier Transform (FFT) is evaluated for each segment.  

The plots on the left of each panel in Fig.~\ref{Fig5} show the evolution of the spectrum as a function of the starting time of the sequence, 
which may be thought of as a time delay relative to $t=0$. The right plot in each panel shows the FFT spectra corresponding to segments 
at the start and end of each simulation. Figures 5(a)-5(d) correspond to parameters which result in thermalization while Fig. 5(e) considers large 
$\lambda$ where the interaction is essentially Caldeira-Leggett, with a weak quadratic nonlinearity, and does not display any signs of thermalization. 
In Fig. 5(f) we show the spectrum of the interspecies interaction energy computed when, within each species, the motion of 
each particle is expressed by a sinusoidal time dependence at its own angular frequency, with equal amplitudes. As noted in the caption,   
the spectrum shows  frequencies of $2\omega_A/2\pi$ and $2\omega_B/2\pi$ which is appropriate given the quadratic dependence of the 
interaction energy on particle separation in the weak coupling limit. This is similar to what is seen in Fig. 5(e) where non thermalization occurs.
In the cases where thermalization occurs, the transfer of spectral strength from the initial frequencies to higher values is clearly inferred 
both from the cascade plots as well as in the two-time FFT plots. 
The cases where we see the evolution of the spectrum to higher frequencies, corresponding to a decay of the original harmonic frequencies, 
occur at smaller $\lambda$ values. By contrast, a considerably more limited spectral evolution is seen in the case (e) where thermalization does not occur. 
This is another clear analogy with fluid mixing where the spatial spectral energy shifts seen there are mimicked by temporal equivalents 
in the dynamics in our setting. This may well be the reason behind the scaling similarities seen both here as well as in our earlier work ~\cite{JauOnoSun2}.

\begin{figure*}[t]
\includegraphics[width=0.49\textwidth, clip=true]{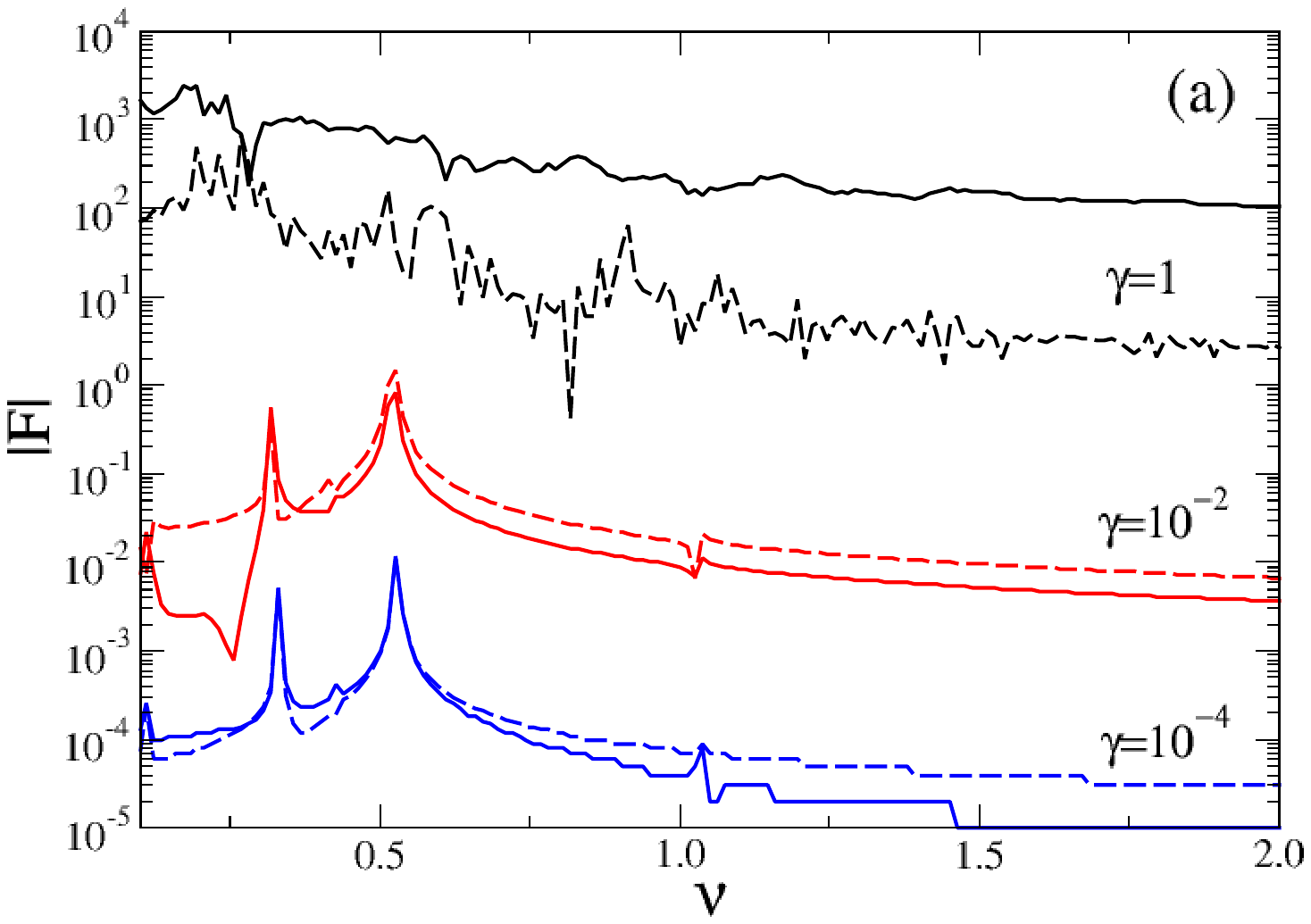}
\includegraphics[width=0.49\textwidth, clip=true]{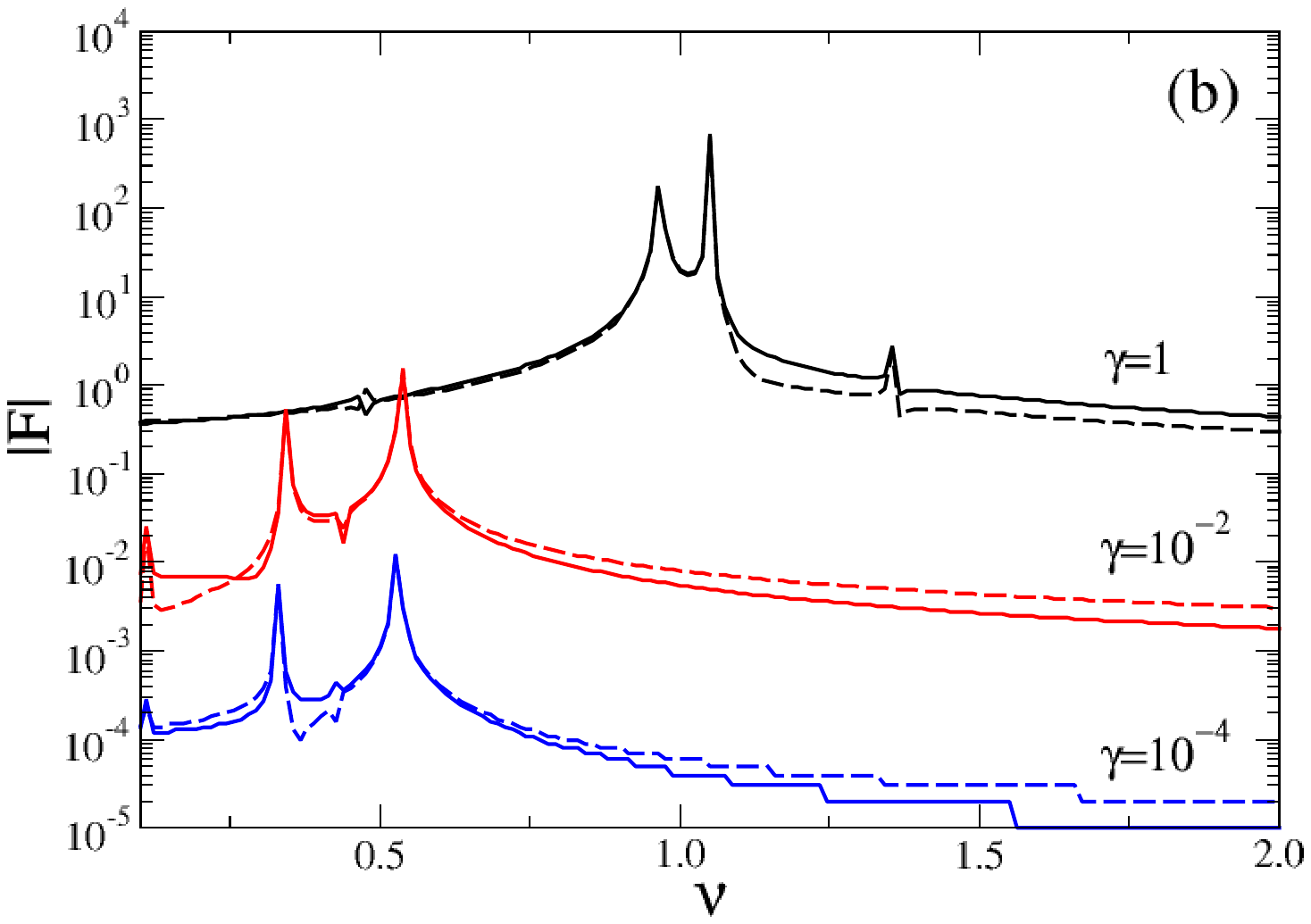}
\caption{Comparison between the Fourier transforms at initial (continuous lines) and final (dashed lines) times in the two cases of (a) the interaction Hamiltonian based on Eq. \ref{Hamilton}, and (b) on its quadratic approximation as in Eq. \ref{Taylor} for $\lambda=10$ and values of $\gamma=10^{-4}$ (blue), $\gamma=10^{-2}$ (red), and 
$\gamma=1$ (black). In these plots we have considered time series of $8192$ time steps in computing the FFT, resulting 
in a higher frequency resolution with respect to the plots in Fig.~\ref{Fig5}.}
\label{Fig6}
\end{figure*}

It should also be noted that, although four cases that result in thermalization are shown, there are underlying differences in 
the mechanism for generating  the 'dynamic bath'. An indication of this feature is visible in the trajectories shown earlier in Fig.~\ref{Fig2}. 
The middle two values of $\lambda=10^{-1}, 1$ correspond to complex, multiparticle dynamics resulting
near the center of the trapping potential while the lowest value of $\lambda=10^{-3}$ corresponds to a very short-range, 
impulsive interaction between particles in the two species. In the latter situation, the spectrum shows the main peak but other 
frequencies do develop in time, leading to the possibility of thermalization. However, these arise from the innate randomness 
in the unequally spaced, episodic 'kicks' when the particles interact.  By contrast, the middle ones show strong frequency 
(harmonic) generation due to the nonlinear nature of the interaction. As already discussed, the largest value of 
$\lambda$ corresponds to a quasilinear, Caldeira-Leggett, regime.

In order to better illustrate the interplay between the frequency spectra and the interaction terms, it is instructive to Taylor 
expand the potential, term felt by the n$^{\mathrm{th}}$ single particle in system A, for instance, in powers of $(q_n-Q_m)/\lambda$, 
that results, modulo a constant

\begin{eqnarray}
&& V_n(q_n, Q_m)= \frac{1}{2} m_A \omega_A^2 q_n^2 +\gamma \sum_{m=1}^{N_B} \exp{[-(q_n-Q_m)^2/\lambda^2]} \nonumber \\ 
& &\simeq \frac{1}{2} m_A \omega_A^2 q_n^2+ \gamma \sum_{m=1}^{N_B} 
\left[1- \left(\frac{q_n-Q_m}{\lambda} \right)^2 + \nonumber \right. \\
& & \left. \frac{1}{2} \left(\frac{q_n-Q_m}{\lambda} \right)^4+{\cal{O}}\left(\frac{q_n-Q_m}{\lambda}\right)^6 +...\right]
\label{Taylor}
\end{eqnarray}

\begin{figure}[t]
\includegraphics[width=0.49\textwidth, clip=true]{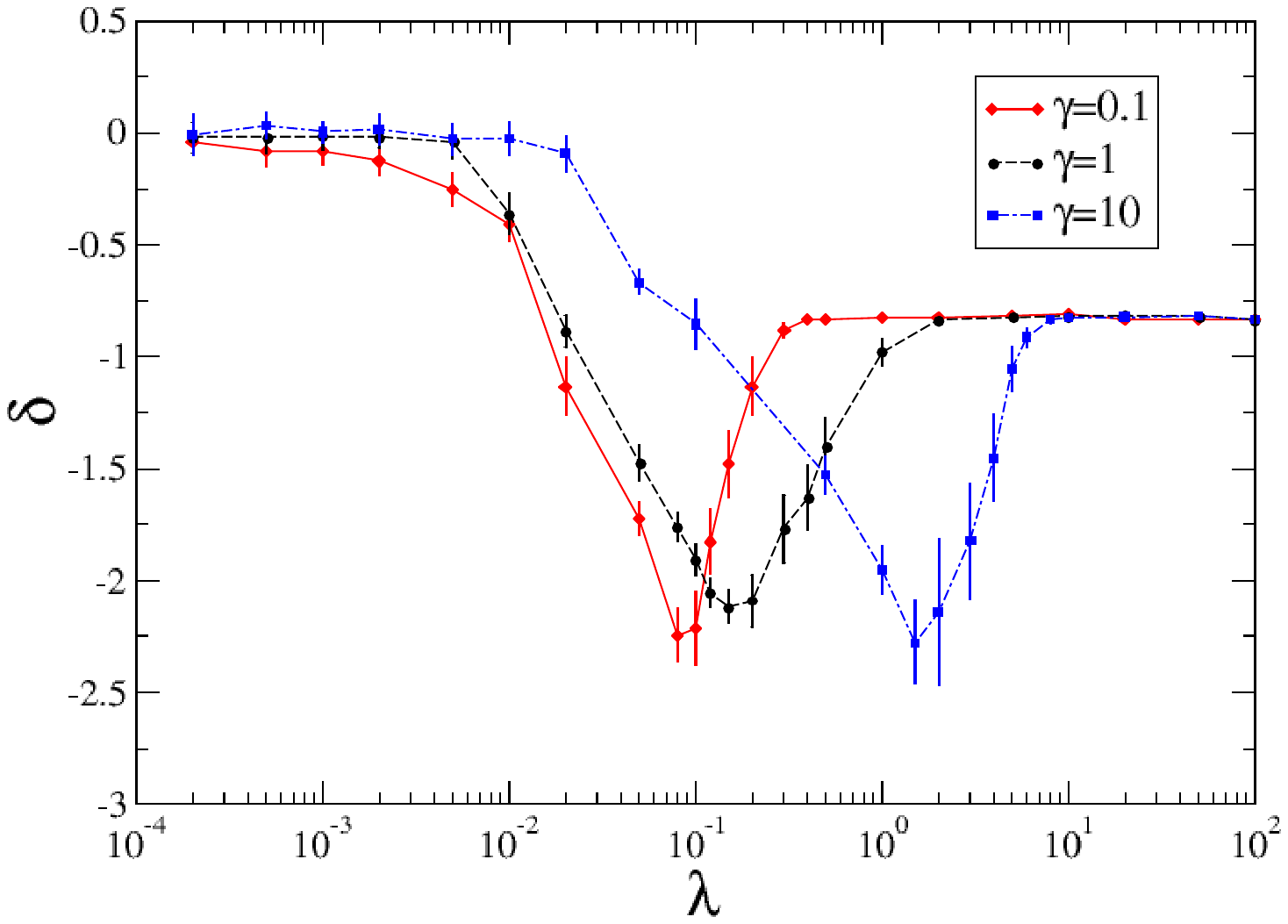}
\caption{Dependence of the scaling exponent $\delta$ upon the interaction range $\lambda$. 
For each value of $\gamma$ and $\lambda$, the FFTs are constructed using the same procedure as indicated in Fig.~\ref{Fig5}, 
using 20 sequences of length 2,048 time steps at the end of the simulation. A range of frequencies is uniformly chosen from 
1/8 to 3/8 of the Nyquist frequency and the FFT values in this range are fitted with a power-law function to extract the exponent 
$\delta$ for each sequence. The 20 values of $\delta$ are then analyzed to get both the mean values and error bars quoted in the figure. 
We have verified that varying the range of frequencies or number of segments considered does not affect the qualitative features of the 
plot  of our analysis. Three regimes are clearly visible corresponding to a nonmonotonic behavior, 
with constant $\delta$ at low $\lambda$, then a sudden decrease until $\delta$ reaches values of about -2, followed by 
an increase at large $\lambda$ to a new plateau corresponding to $\delta \simeq -0.8$. 
The data also show the dependence for three distinct values of $\gamma$, most notably an increase of the 
value of $\lambda$, while increasing $\gamma$, for which $\delta$ reaches its minimum value.} 
\label{Fig7}
\end{figure}

The quadratic terms in $q_n$ can be rearranged to provide a renormalized angular frequency $\omega_A'$ 

\begin{equation}
\omega_A'= \omega_A \sqrt{1 - {\frac{\gamma N_B}{m \omega_A^2 \lambda^2}}},
\label{renorm}
\end{equation}
while the same oscillator interacts linearly with all the oscillators of system B. Obviously these considerations also hold for 
any single oscillator of system B. This results, even without considering the quartic term and higher order terms, in $N_B+1$ normal 
modes with nondegenerate frequencies. Therefore we see that already at this level, {\it i.e.} for $\lambda$ not too small to 
make the Taylor expansion invalid, other frequencies contribute to the Fourier transform though only in a narrow band. 
This is further enhanced and broadened by the presence of the higher order terms resulting in frequency mixing and harmonic generation.
The spectral broadening can be associated with chaotic dynamics, as expected even at the quartic term level 
\cite{Lakshmanan,Lakshminarayan,Bannur}. Notice that, while the case of $\gamma<0$ in Eq.~\ref{renorm} just adds stiffness 
to the preexisting harmonic potential, the repulsive case of $\gamma>0$ may originate a region of antitrapping, which 
eventually may lead to a situation quite similar to the one of the Sinai model \cite{Sinai} in which hard spheres give 
rise to a nonintegrable system. Then we expect less intriguing behavior in the attractive case, and for this reason it  
has not been considered in this contribution. In Fig.~\ref{Fig6}, we show a comparison between the FFTs of the 
initial and final segments taken from a long interaction energy series for systems evolving as in Eq. \ref{Hamilton} (left panel)
contrasted with evolution under the leading, quadratic order term in the Taylor expansion in Eq.~\ref{Taylor} (right panel). 
Notice that for the strong coupling case of $\gamma=1$ the FFT peaks are shifted at higher frequencies for the leading order 
 approximation, as expected from Eq.~\ref{Taylor}. At large $\gamma$ the negative curvature due to the coupling to the oscillators of the 
 other bath prevails and the effective potential becomes bistable, a phenomenon already discussed in detail in \cite{OnoSun}, 
 and also visible in panel (c) of Fig.~\ref{Fig2} with the particle moving, at later times, around two symmetric locations.
In this approximation we expect small oscillations of the position of each oscillator around one of the two minima of 
 the bistable potential at the angular frequency $\Omega_A$
 
 \begin{equation}
\Omega_A= \omega_A \sqrt{2\left(\frac{\gamma N_B}{m \omega_A^2 \lambda^2}-1\right)},
\label{renormbistable}
\end{equation}

 The predicted peaks for the Fourier transform of the interaction energy, based on this approximation, occur at frequencies $2 \Omega_A/(2\pi)  \simeq 0.78$ 
 and $2 \Omega_B/(2 \pi) \simeq 1.35$, in arbitrary units, to be compared to the observed ones  of $0.96$ and $1.04$, respectively. 
 Even within the quadratic approximation, there is some limited broadening due to the multiplicity of normal modes, as already 
 discussed earlier and, in the strong coupling regime, also due to the nonlinearity resulting from the effective bistable potential. 
 The left panel of Fig.~\ref{Fig6} also shows the dramatic effect of the genuine nonlinearities induced by the full interaction energy.
More specifically, further low frequency components are enhaced at the final times for large $\gamma$, as expected for the onset 
of thermalization. Such an enhancement is not visible at small $\gamma$, and correspondingly there is no thermalization on the 
time scale of the simulation. Moreover, for the full Gaussian interaction the peaks are no longer well defined at large $\gamma$, and 
the Fourier spectrum is basically featureless even at early times.
 These insights are also corroborated in Fig.~\ref{Fig1}, where the case of nondegenerate angular frequencies result in a
 broader anomalous region and faster  thermalization when compared to the degenerate case. All these considerations 
 are affected by the final temperature of the baths which tends to wash out the Fourier transform, as shown in Fig. 7 of earlier
 work~\cite{Choi} discussing a bistable potential interacting with a Caldeira-Leggett bath.
 
The distinction between the various regimes identified in the discussion of Figs.~\ref{Fig5} and \ref{Fig6} becomes clearer with 
further quantitative considerations, by focusing on the high-frequency regime of the Fourier spectra at late times.
These tails are adequately fitted with a power-law dependence, $S(\omega) \simeq \omega^{\delta}$. The outcome of this analysis is shown
in Fig.~\ref{Fig7} versus $\lambda$ and for three values of the interaction strength $\gamma$. At low $\lambda$ the spectrum is flat, compatible with 
$\delta =0$, which is consistent with the sudden exchange of energy during the rare collisions, as already remarked in interpreting Fig.~ \ref{Fig2}. 
In the opposite limit of large $\lambda$ the absence of mixing due to the very weak nonlinearity implies that the two original trapping frequencies 
are preserved. For such a system with contributions to the Fourier spectrum from two frequencies, as verified independently from panel (f) 
of Fig.~\ref{Fig5}, we expect $\delta \simeq 0.8$, which is in line with the accurate determinations of $\delta$ at the largest $\lambda$. 
In between, the nonlinearities generate several harmonics nearby the two original frequencies, generating a cascade of harmonics such as 
$|\omega_A \pm \omega_B|$ and all multiples. We therefore expect that most of the spectral density is concentrated around the two frequencies. 
The expected Kolmogorov scaling makes sure that at the smallest timescales this intermediate regime should have depleted Fourier 
components, although currently we cannot provide an immediate quantification of this effect. Nevertheless, the parameter $\delta$ appears a more 
reliable indicator for the presence of an intermediate regime between the purely Boltzmann case at small $\lambda$ and the Caldeira-Leggett case 
at very large $\lambda$, as seen by comparing the quite scattered plots in Fig.~\ref{Fig1} and the clean structures present in Fig.~\ref{Fig7}. 
The minimum value of $\delta$ is approximately constant, and shifted to larger values of $\lambda$ as the interaction strength is increased, suggesting 
a combined dependence on $\gamma$ and $\lambda$.  

\begin{figure}[t]
\includegraphics[width=0.49\textwidth, clip=true]{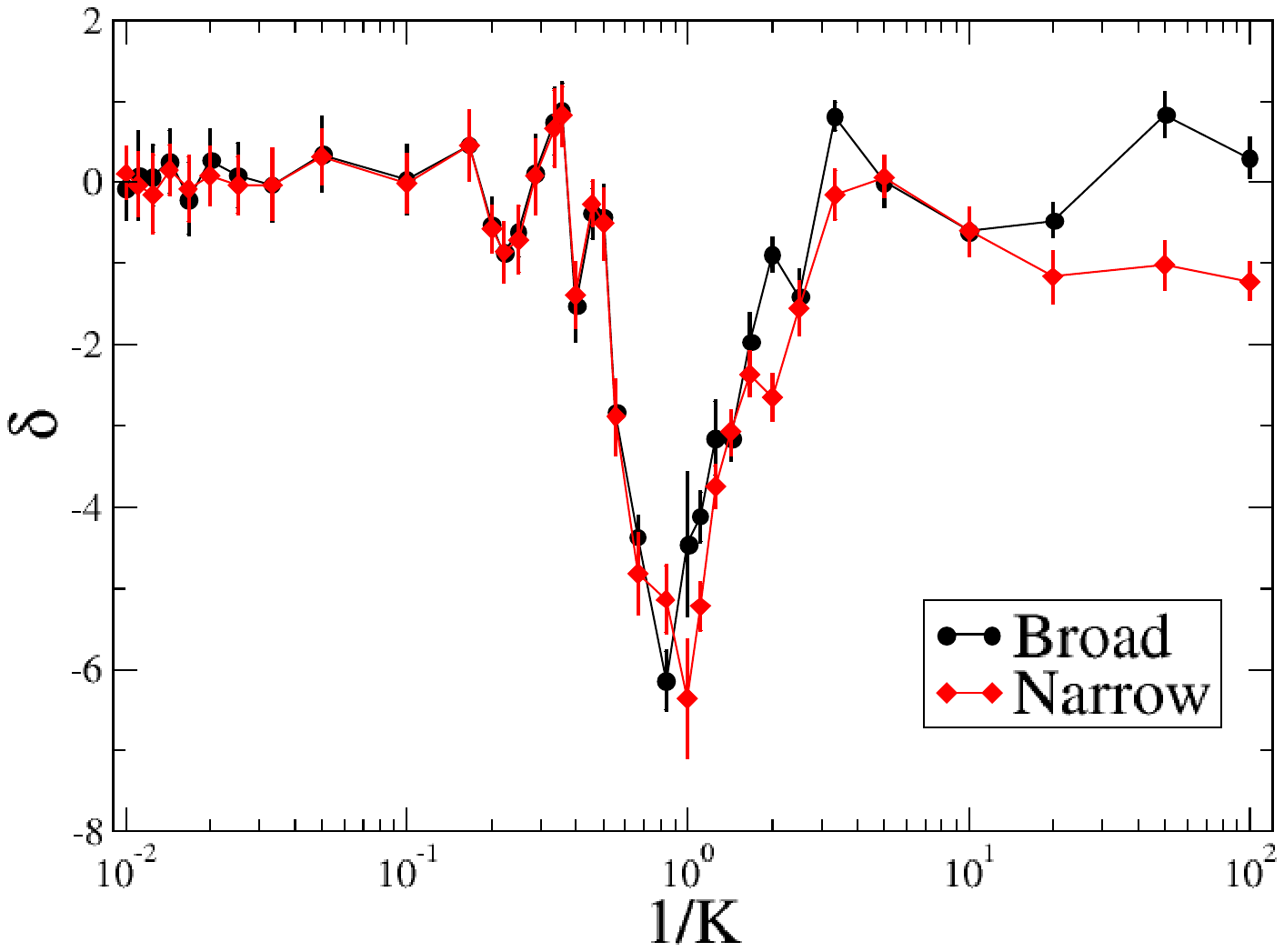}
\caption{Dependence of the Fourier scaling exponent $\delta$ for the simple standard map dynamics upon the 
stochasticity parameter $K$. For each $K$, a long time series ($10^5$ time steps) of the momentum is generated for a set of initial
$40$ conditions. The ensemble averaged momentum is used for constructing the FFTs and an analogous procedure as  adopted 
for generating Fig.~\ref{Fig7} is followed to extract the exponent $\delta$ and its error bars for each sequence. 
These values are plotted against $1/K$ tto allow for a better inferential match with the case of the Gaussian interaction potential, 
since the role of $K$ in the latter system is played by $\gamma/\lambda^2$. 
Note that large $K$ corresponds to global chaos in the standard map while for values $1 \lesssim K \lesssim 4$ stable regions coexist with chaos. 
The two curves labelled broad (circles) and narrow (diamonds) correspond to initial conditions drawn from the entire domain and from a limited 
(10\% of the allowed domain) area centered at a randomly selected point in phase space. Beyond the threshold for global chaos, the
behavior is independent of the choice.}
\label{Fig8}
\end{figure}

From a dynamical systems perspective, spectral signatures of chaotic dynamics in the form of broadening have been investigated for many
years~\cite{LinsayFFT,Abarbanel}. Given that the exponent $\delta$ shows well-defined behavior in Fig.~\ref{Fig7}, as the interaction range
of the Gaussian potential is varied, this would appear to a context for exploring the role of chaotic dynamics in the system. Given the inherently
high dimensionality of our classical phase space, it s worth exploring if a paradigmatic, low-dimensional dynamical system, such as the 
standard map~\cite{LichtLieb} exhibits similar trends. This would provide an indication of the parameter regimes, in our problem, where 
the presence of chaos assists in the thermalization process.

With this as motivation, we consider the well-studied Chirikov-Taylor or standard map, an area preserving mapping resulting from a periodically
kicked Hamiltonian 
\begin{equation}
H(p,q,t)=p^2/2 + K\sin{q} \sum_n \delta(t-n)\;,
\end{equation}
where spacing between kicks has been set to $1$. The resulting dynamics are governed by the $2D$ mapping:
\begin{eqnarray}
p_{n+1} &=& p_{n} + K \cos{q_n} \nonumber \\
q_{n+1} &=& q_n + p_{n+1}\;,
\end{eqnarray}
where both $p,q$ will be considered modulo $2\pi$.
The kick strength $K$ acts as the stochasticity parameter that determines the character of the dynamics. 
For small $K$ values, below a critical threshold $K_c = 0.9716\dots$ , the increase in momentum is bounded by 
local regions separated by invariant curves and any chaotic dynamics is local~\cite{LichtLieb}. 
On exceeding $K_c$, the last demarcating boundary disappears and all regions of phase space become accessible.  
The resulting phase space is mixed in terms of the dynamics with stable, regions embedded in a chaotic background. 
These stable regions shrink with further increase in $K$. Beyond $K=4$, the phase space is dominantly chaotic with very small,
isolated regions of stable dynamics which can reappear for specific parameter windows. For the purposes of our discussion, the
essential feature is the chaos dominated dynamics for $K>4$.

We use the map to generate a long time series of $100,000$ points for a set ($40$) of initial conditions and,
after computing a time series of the ensemble averaged momentum,
follow an analogous protocol to that used in generating Fig.7. Namely, several segments of 
$2,048$ points were chosen at later times, the FFT constructed and a range of frequencies fitted
to find the exponent $\delta$. The window considered has to be adjusted in the case of the standard
map due to an important difference in the dynamics. In the case of the nonlinear, Gaussian interaction
potential, harmonic generation results in higher frequencies being generated in the dynamics. By contrast, 
for the standard map, as the highest frequency is set by the time between kicks, sub-harmonics are generated by the
dynamics, rather than harmonics. Also, at the smaller values of the parameter $K$, the existence
of isolated peaks in the frequency spectrum  makes this fitting suspect. However, as stated earlier, our main
interest is in the regime where chaos dominates the dynamics and, here, the method works well. We consider
a wide range of $K$ values and, as seen from Fig.~\ref{Fig8}, it is more useful plotting $\delta$ versus $1/K$ as this 
better aligns with the results shown in Fig.~\ref{Fig7}.

We also considered two versions of the initial condition ensemble, where the first (labeled 'Broad') was chosen
over the full extent of the allowed space while the second ('Narrow') was taken from a narrow region centered at
a randomly picked point in phase space. For lower values of $K$, local structures in phase space influence $\delta$
and this is reflected in the differences between the two ensembles. This inference is readily supported by considering
the same analysis for single initial conditions. However, once the chaotic dynamics becomes global, the trends in
$\delta$ become largely independent of the initial ensemble. 

The principal inference to be drawn from this exercise is that the spectrum is largely flat ($\delta \approx 0$) at large
values of $K$ where the dynamics is dominated by chaos. This is what is observed in Fig.~\ref{Fig7} for the Gaussian
potential at smaller $\lambda$ and for larger values of the interaction strength $\gamma$. We also have the limiting
description in the limit $\lambda \rightarrow 0$ where the random, hard-sphere interactions are akin to a Sinai
billiards where the dynamics is known to be chaotic. Overall, these results would appear to support the inference
that there is a parameter range in our problem where chaos, arising from the nonlinearities in the interaction potential,
assists in thermalization.  Further discussions on the chaos to thermalization connection in a variety of dynamical 
systems with small number of degrees of freedom can be found in \cite{Wilkinson,Berry,Jarzynski,Bonanca,Rosa}. 
Taken together with the earlier discussions, our results suggest the existence of different mechanisms that promote 
thermalization (or lack thereof) as the range of the interaction potential is varied.

\section{Conclusions}

We have discussed the thermalization process for two species harmonically trapped in the presence of a nonlinear interspecies potential of 
Gaussian form. This potential interpolates the two extreme regimes of rare, local collisions, as in the Boltzmann approach, and the case of linear 
coupling between harmonic oscillators characteristic of the Caldeira-Leggett approach. This analysis is somewhat complementary to the one 
presented in \cite{Morrison} where a system described by the Vlasov-Poisson equation, for instance a collisionless plasma, has been mapped 
into a Caldeira-Leggett-like setting. In an intermediate regime of a characteristic interaction length, there is the regime of frequent, weak interactions 
which is usually captured by the physics of the Fokker-Planck equation. It is worth remarking that this model connects these three different approaches
in a unique scenario involving only interaction strength and range as free parameters. In principle there are no limitations on exploring arbitrary large 
coupling strengths far from the weak, perturbative regime. However, we do not necessarily expect Boltzmann energy distributions in the strong coupling 
regime, which will require a careful future analysis and the need to introduce new parameters replacing the concept of temperature. 

The regime of intermediate ranges exhibits anomalies for the behavior of total interaction energy at equilibrium. 
This is manifested both as an anomalous exponent in the dependence of this  quantity upon the number of involved particles, as 
well as through a non-monotonic behavior of the exponent ruling the power-law dependence of the its Fourier transform in the high-frequency tail.
This second indicator seems quite sensitive to the anomalous behavior, and allows to characterize the thermalization stage via the spectral study 
of the interaction energy, a viewpoint recently discussed, in the context of open quantum systems, in \cite{Xu}.
Further, the analogous behavior of the exponent in our Gaussian model and the simple standard map suggest that
chaotic dynamics resulting from nonlinearities in the interaction potential may be involved in the thermalization process for certain parameter regimes.

The non-extensive feature of the interaction energy at intermediate ranges is in line with what is expected in models defined by 
non-extensive variables relevant for the description of statistical systems with medium and long-range interactions \cite{Tsallis}.  
It is already known, from detailed studies of the Hamiltonian Mean Field (HMF) model \cite{Konishi,Ruffo}, that long-range 
interactions in a many body system result in deviations from Maxwell-Boltzmann energy distribution in the form of $q$-distributions, 
ergodicity breaking, and deviations from the central limit theorem for dynamical variables \cite{Baldovin,Rapisarda,Pluchino}. 
In this framework, our model may provide a simple setting for a many body system fully defined in phase space, with weak or 
strong correlations depending on the coupling strength $\gamma$, and complementary to the HMF model.
In particular,  in the future we aim at a careful study of the energy distribution to look for deviations, at intermediate 
or long times, from the initially imprinted Boltzmann distribution. This could complement in our model recent findings on nonlinear
 models such as the Fermi-Pasta-Ulam-Tsingou model \cite{BagchiTsallis}, with predictions of anomalous diffusion \cite{Plastino,Marin}, and 
 analysis of long-range interactions \cite{PlastinoPlastino,Escamilla}. Another issue worth considering will be the exploration 
 of regimes where the results presented here have a bearing on quantum or, at least, semi-classical approaches to many-body thermalization.


\begin{thebibliography}{99}

\bibitem{Maxwell} J. C. Maxwell, 
{\sl IV. On the Dynamical Theory of Gases},
Phil. Trans. R. Soc. Lond. A \textbf{157}, 49 (1867). 

\bibitem{Boltzmann} L. Boltzmann, {\sl Lectures on Gas Theory} (Dover, New York, 1995), reprint of the 1986-1898 edition.

\bibitem{Cercignani} C. Cercignani, {\sl The Boltzmann Equation and Its Applications} (Springer-Verlag, New York, 1988).

\bibitem{Kremer} G. M. Kremer, {\sl An Introduction to the Boltzmann Equation and Transport Processes in Gases} (Springer, Berlin, 2010).

\bibitem{Magalinskii} V. B. Magalinskii, 
Dynamical model in the theory of the Brownian motion,
Sov. Phys. JETP \textbf{9}, 1382 (1959).

\bibitem{Ullersma1} P. Ullersma,
An exactly solvable model for Brownian motion: 1. Derivation of Langevin equation,
Physica \textbf{32}, 27 (1966).

\bibitem{Ullersma2} P. Ullersma,
An exactly solvable model for Brownian motion: 2. Derivation of Fokker-Planck equation and master equation,
Physica \textbf{32}, 56 (1966).

\bibitem{Ullersma3} P. Ullersma,
An exactly solvable model for Brownian motion: 3. Motion of a heavy mass in a linear chain,
Physica \textbf{32}, 74 (1966).

\bibitem{Ullersma4} P. Ullersma,
An exactly solvable model for Brownian motion: 4. Susceptibility and Nyquist's theorem,
Physica \textbf{32}, 90 (1966).

\bibitem{Caldeira1} A. O. Caldeira and A. J. Leggett, 
Influence of dissipation on quantum tunneling in macroscopic systems,
Phys. Rev. Lett. \textbf{46}, 211 (1981).

\bibitem{Caldeira2} A. O. Caldeira and A. J. Leggett,
Quantum tunneling in a dissipative system,
Ann. Phys. \textbf{149}, 374 (1983).

\bibitem{Caldeira3} A. O. Caldeira and A. J. Leggett,
Influence of damping on quantum interference: an exactly soluble model,
Phys. Rev. A \textbf{31}, 1059 (1985).

\bibitem{OnoSun} R. Onofrio and B. Sundaram, 
Effective microscopic models for sympathetic cooling of atomic gases,
Phys. Rev. A \textbf{92}, 033422 (2015).

\bibitem{JauOnoSun1} F. Jauffred, R. Onofrio, and B. Sundaram, 
Universal and anomalous behavior in the thermalization of strongly interacting harmonically trapped gas mixtures, 
J. Phys. B: At. Mol. Opt. Phys \textbf{50}, 135005 (2017).

\bibitem{Kolmogorov} A. N. Kolmogorov, 
The local structure of turbulence in incompressible viscous fluid for very large Reynolds numbers,
Dokl. Akad. Nauk SSSR \textbf{30}, 299 (1941).

\bibitem{JauOnoSun2} F. Jauffred, R. Onofrio, and B. Sundaram, 
Simulating sympathetic cooling of atomic mixtures in nonlinear traps,
Phys. Lett. A \textbf{381}, 2783 (2017).

\bibitem{JauOnoSun3} F. Jauffred, R. Onofrio, and B. Sundaram, 
Scaling laws for harmonically trapped two-species mixtures at thermal equilibrium,
Phys. Rev. E \textbf{99}, 022116 (2019).

\bibitem{Mensky} A. Konetchnyi, M. Mensky, and V. Namiot, 
Physical model for monitoring the position of a quantum particle,
Phys. Lett. A \textbf{177}, 283 (1993).

\bibitem{LichtLieb}  See, for example, A. J. Lichtenberg and M. A. Lieberman, {\sl Regular and Chaotic Dynamics}, AMS vol. 38 (Springer, 1992).

\bibitem{Smith} S. T. Smith and R. Onofrio, 
Thermalization in open classical systems with finite heat baths, 
Eur. Phys. J. B \textbf{61}, 271 (2008).

\bibitem{Lakshmanan} M. Lakshmanan and R. Sahadevan, 
Painlev\'e analysis, Lie simmetries, and integrability of coupled nonlinear 
oscillators of polynomial type,
Phys. Rep. \textbf{224}, 1 (1993).
 
\bibitem{Lakshminarayan} A. Lakshminarayan, M. S. Santhanam, and V. B. Sheorey, 
Local scaling in homogeneous Hamiltonian systems,
Phys. Rev. Lett. \textbf{76}, 396 (1996).

\bibitem{Bannur} V. M. Bannur, P. K. Law, and J. C. Parikh,
Statistical mechanics of quartic oscillators,
Phys. Rev. E \textbf{55}, 2525 (1997).

\bibitem{Sinai} Ya. G. Sinai, 
On the foundations of the ergodic hypothesis for a dynamical system of statistical mechanics,
Dokl. Akad. Nauk SSSR \textbf{153}, 1261 (1963).

\bibitem{Choi} S. Choi, R. Onofrio, and B. Sundaram, 
Ehrenfest approach to open double-well dynamics,
Phys. Rev. E \textbf{92}, 042907 (2015).

\bibitem{LinsayFFT} See for example P. S. Linsay, Period Doubling and Chaotic Behavior in a Driven,
Anharmonic Oscillator, \prl \textbf{47}, 1349 (1981).

\bibitem{Abarbanel} H. Abarbanel, {\sl Analysis of Observed Chaotic Data}, (Springer, 1995).

\bibitem{Wilkinson} M. Wilkinson,
Dissipation by identical oscillators,
J. Phys. A: Math. Gen. \textbf{23}, 3603 (1990).

\bibitem{Berry} M. V. Berry and J. M. Robbins, 
Chaotic classical and half-classical adiabatic reactions: geometric magnetism and deterministic friction,
Proc. R. Soc. Lond. A \textbf{442}, 659 (1993).

\bibitem{Jarzynski} C. Jarzynski, 
Thermalization of a Brownian particle via coupling to low-dimensional chaos,
Phys. Rev. Lett. \textbf{74}, 2937 (1995).

\bibitem{Bonanca} M. V. S. Bonan\c{c}a and M. A. M. de Aguiar,
Classical dissipation and asymptotic equilibrium via interaction with chaotic systems,
Physica A \textbf{365}, 333 (2006).

\bibitem{Rosa} J. Rosa and M. W. Beims, 
Dissipation and transport dynamics in a ratchet coupled to a discrete bath,
Phys. Rev. E \textbf{78}, 031126 (2008).

\bibitem{Morrison} G. I. Hangstrom and P. J. Morrison, 
Caldeira-Leggett model, Landau damping, and the Vlasov-Poisson system,
Physica D \textbf{240}, 1652 (2011).

\bibitem{Xu} S. Xu, H. Z. Shen, X. X. Yi, and W. Wang, 
Readout of the spectral density of an environment from the dynamics of an open system,
Phys. Rev. A \textbf{100}, 032108 (2019).

\bibitem{Tsallis} C. Tsallis, {\sl Introduction to Nonextensive Statistical Mechanics}, vol. 34 (Springer, 2009).

\bibitem{Konishi} T. Konishi and K. Kanenko, 
Clustered motion in symplectic couple map systems,
J. Phys. A \textbf{25}, 6283 (1992).

\bibitem{Ruffo} M. Antoni and S. Ruffo, 
Clustering and relaxation in Hamiltonian long-range dynamics,
Phys. Rev. E \textbf{52}, 2361 (1995).

\bibitem{Baldovin} F. Baldovin and A. L. Stella, 
Central limit theorem for anomalous scaling due to correlations,
Phys. Rev. E \textbf{75}, 020101(R) (2007).

\bibitem{Rapisarda} C. Tsallis, A. Rapisarda, A. Pluchino, E. P. Borges, 
On the non-Boltzmannian nature of quasi-stationary states in long-range interacting systems,
Physica A \textbf{381}, 143 (2007).

\bibitem{Pluchino} A. Pluchino, A. Rapisarda, and C. Tsallis,
Nonergodicity and central-limit behavior for long-range Hamiltonians,
EPL \textbf{80}, 26002 (2007).

\bibitem{BagchiTsallis} D. Bagchi and C. Tsallis, 
Long-ranged Fermi-Pasta-Ulam systems in thermal contact: Crossover from $q$-statistics to Boltzmann-Gibbs statistics,
Phys. Lett. A \textbf{381}, 1123 (2017).

\bibitem{Plastino} A. R. Plastino and A. Plastino, 
Non-extensive statistical mechanics and generalized Fokker-Planck equation,
Physica A \textbf{222}, 347 (1995).

\bibitem{Marin} D. Marin, M. A. Ribeiro, H. V. Ribeiro, and E. K. Lenzi, 
A nonlinear Fokker-Planck equation approach for interacting systems: Anomalous diffusion and Tsallis statistics,
Phys. Lett. A \textbf{382}, 1903 (2018).

\bibitem{PlastinoPlastino} A. R. Plastino and A. Plastino, 
Stellar polytropes and Tsallis's entropy,
Phys. Lett. A \textbf{174}, 384 (1993).

\bibitem{Escamilla} L. F. Escamilla-Herrera, C. Gruber, V. Pineda-Reyes, and H. Quevedo, 
Statistical mechanics of the self-gravitating gas in the Tsallis framework,
Phys. Rev. E \textbf{99}, 022108 (2019).

\end{thebibliography}
\end{document}